\documentclass[letterpaper,twocolumn]{article}

\usepackage[T1]{fontenc}

\usepackage{geometry}
\usepackage{setspace}

\usepackage[backend=biber,style = chem-acs,articletitle=true]{biblatex}
\usepackage{graphicx}
\usepackage{float}

\usepackage{color}
\usepackage[usenames,dvipsnames]{xcolor}
\usepackage{hyperref}
\hypersetup{colorlinks=true,linkcolor=blue,urlcolor=blue,citecolor=blue}

\usepackage{authblk}
\author[1,2]{Alexey A. Sokolik*}
\author[3,2]{Azat F. Aminov}
\author[3]{Evgenii E. Vdovin}
\author[3]{Yurii N. Khanin}
\author[4,5]{Mikhail A. Kashchenko}
\author[6,7]{Denis A. Bandurin}
\author[3]{Sergey V. Morozov}
\author[7]{Kostya S. Novoselov}

\affil[1]{Institute for Spectroscopy, Russian Academy of Sciences, 108840 Troitsk, Moscow, Russia}
\affil[2]{National Research University Higher School of Economics, 109028 Moscow, Russia}
\affil[3]{Institute of Microelectronics Technology and High Purity Materials, Russian Academy of Sciences, 142432 Chernogolovka, Russia}
\affil[4]{Programmable Functional Materials Lab, Center for Neurophysics and Neuromorphic Technologies, 127495 Moscow, Russia}
\affil[5]{Moscow Center for Advanced Studies, 123592 Moscow, Russia}
\affil[6]{Department of Materials Science and Engineering, National University of Singapore, 117575 Singapore}
\affil[7]{Institute for Functional Intelligent Materials, National University of Singapore, 117544 Singapore}

\title{Tunneling characteristics of twisted double bilayer graphene heterostructures}
\date{*Email: asokolik@hse.ru}

\begin{document}

\maketitle

\begin{abstract}
Electron tunneling between sheets of bilayer Bernal graphene twisted at different small angles was studied experimentally and theoretically. The current-voltage characteristics exhibit resonant peaks, steps, and regions of negative differential resistance, the origin of which is explained by the intersections of energy- and momentum-shifted electron dispersions of adjacent layers. A theoretical analysis of tunneling transport demonstrated that the key to understanding this phenomenon lies in the competition between two contributions: between like (conductivity-conductivity or valence-valence) and unlike (conductivity-valence) bands of parallel bilayer graphene sheets. A systematic evolution of the tunneling current patterns with increase of the twist angle is investigated. Polarization of electron wave function across graphene sublayers caused by displacement field within bilayer graphene is shown to strongly affect the tunneling probability, thus enhancing negative differential resistance due to Van Hove singularities at the band edges. 
\end{abstract}

\section*{Keywords}

bilayer graphene, tunnel junctions, negative differential resistance, van der Waals heterostructures

\section*{}

\begin{figure}[t]
\centering
\includegraphics[width=0.45\textwidth]{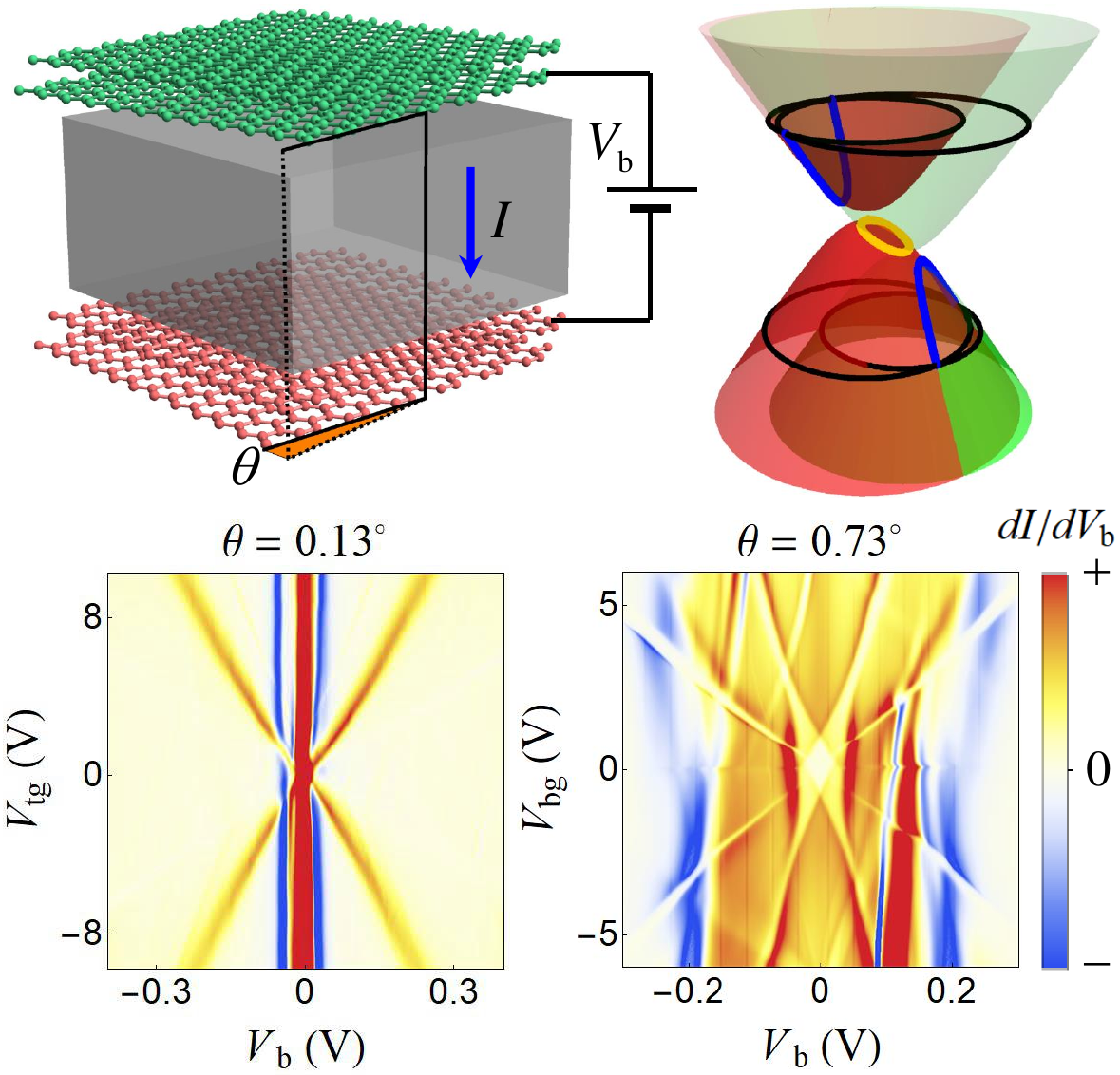}
\end{figure}

Tunneling of electrons in van der Waals heterostructures has become an important area of research in recent years. Unlike traditional epitaxial tunneling structures based on semiconductors, the van der Waals heterostructures combine atomically thin layers of conducting and isolating materials with possibility to finely control relative crystallographic orientation of adjacent layers \cite{Geim2013,Novoselov2016}. Such structures allowed to study a series of new collective quantum effects such as Coulomb drag \cite{Kim2011,Gorbachev2012}, interlayer excitons \cite{Jiang2021,Regan2022}, dipolar superfluidity and enhanced tunneling due to interlayer exciton condensation \cite{Li2017,Burg2018,Nguyen2025}, and other phenomena including photocurrent generation and photoluminescence \cite{Shehabeldin2026}.

Considerable effort has been devoted to the study of negative differential resistance on I-V characteristics of the tunneling devices originally made from monolayer graphene sheets \cite{Britnell2013,Mishchenko2014,Zhang2023,Zhang2025,Zhang2026,Kuzmina2021}, which is promising from the point of view of applied electronics. The structures based on Bernal bilayer graphene (BLG) differ by quadratic electron dispersion, opening of the gap in vertical electric field \cite{McCann2013}, and redistribution of electron wave functions over graphene sublayers affecting tunneling current \cite{Vdovin2024,Sokolik2025} and promoting its valley polarization \cite{Thompson2019}. The regions of negative differential resistance in bilayer graphene heterostructures were first discovered in Refs.~\cite{Fallahazad2015,Kim2016,Burg2017}, where theoretical modeling of I-V characteristics was also carried out, and the peaks of tunneling current were explained in terms of intersection of electron dispersions in different sheets \cite{delaBarrera2015,Zhang2025}. More recent studies revealed the role of finite lifetime and parallel magnetic field \cite{Prasad2021}, and possible formation of interlayer excitons \cite{Burg2018}. However these studies were restricted to heterostructures without relative twist of BLG sheets, and gap opening in BLG due to vertical electric field was not taken into account in the calculations.

\begin{figure}[t]
\centering
\includegraphics[width=\columnwidth]{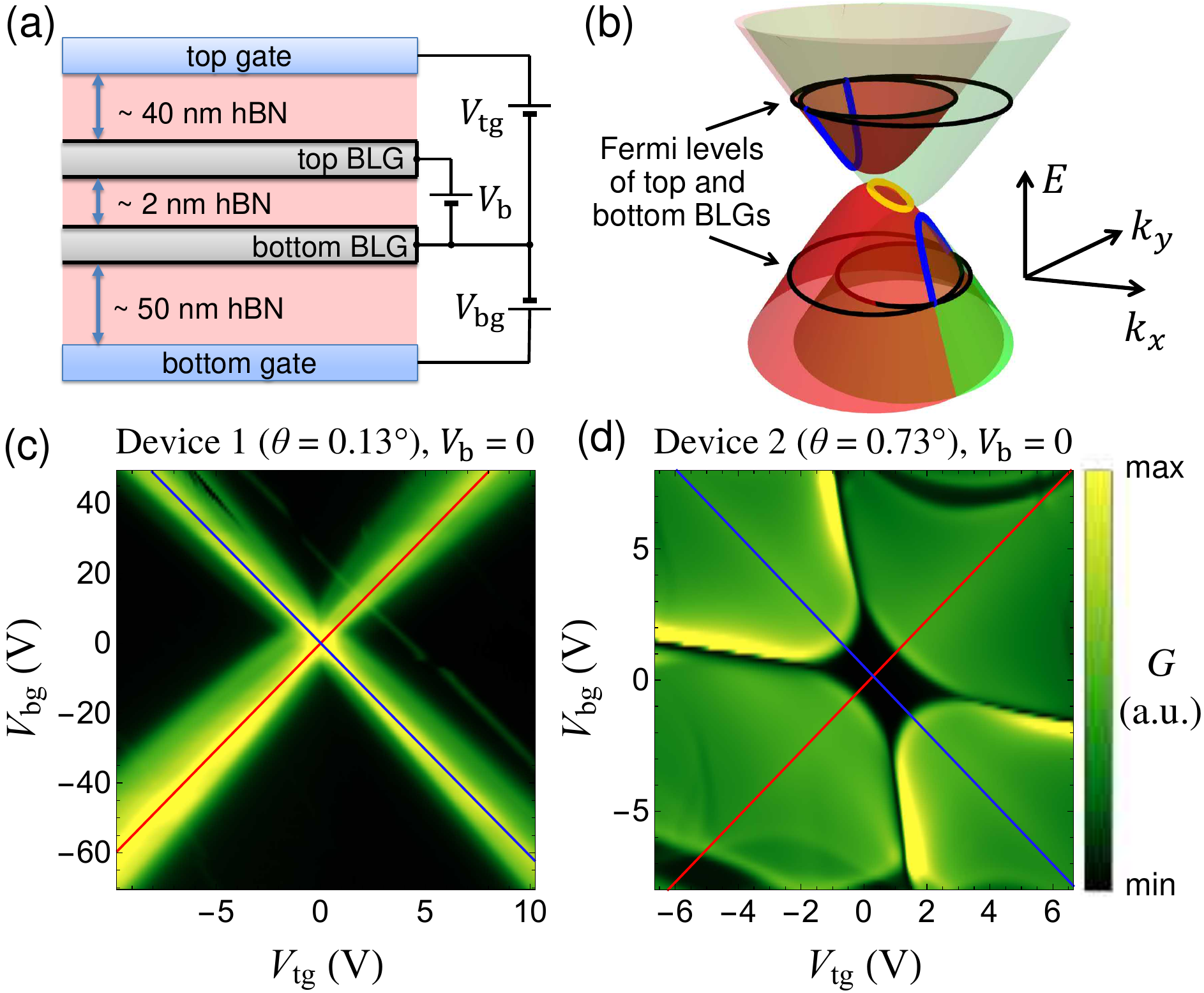}
\caption{Electron tunneling between mutually twisted BLGs in dual-gated system. (a) System schematic for Device 2; Device 1 has additional 290-nm thick $\mathrm{SiO}_2$ layer above the bottom gate. (b) Electron dispersions in top (red) and bottom (green) BLG, relatively shifted in momentum and energy, which exhibit like- (blue curves) and unlike-band (yellow curve) intersections. The parts of electron dispersions corresponding to unoccupied states are semitransparent. (c,d) Zero-bias tunneling conductance as function of top ($V_\mathrm{tg}$) and bottom ($V_\mathrm{bg}$) gate voltages in Device 1 and Device 2, respectively. Red and blue lines show equal- and opposite-density diagonals, respectively. Black stripes in the corners of (d) are likely caused by a secondary Dirac point due to Moire potential at the hBN/BLG interface \cite{Sokolik2025} and lie beyond the scope of our study.}
\label{Fig1}
\end{figure}

Our paper presents experimental and theoretical study of the finite-bias tunneling in twisted heterostructures between Bernal BLG sheets with different twist angles. High homogeneity of the samples and ultrathin layers allows us to take into account only the tunneling transitions with conservation of lateral electron momentum and energy. As a consequence, the tunneling occurs through intersections of electron dispersion surfaces in two spatially separated BLG sheets, which are displaced with respect to each other in energy (due to the bias voltage) and in momentum space (since crystal lattices are relatively twisted). An example of such intersections is shown in Figure~\ref{Fig1}(b). 

The dispersion intersections from like (conduction-conduction or valence-valence) and unlike  (conduction-valence) bands of BLGs provides two channels for the tunneling current, which forms the basis for theoretical analysis of the measured I-V characteristics. Comparison of the calculation results with the experimental data reveals important role of the twist angle, as well as sublayer polarization of van Hove singularities arising when a gap is opened in either BLG. In addition, we highlight the existence of two qualitatively different types of I-V characteristics --- symmetric and asymmetric --- depending on gate voltages, and analyze their main features: peaks, steps, and regions of negative differential resistance, which are reproduced by the theoretical calculations with good accuracy. 

Two devices with different twist angles between BLG sheets were studied in experiment: $\theta=0.13^\circ$ for Device 1 and $\theta=0.73^\circ$ for Device 2. Both devices have dual-gate architecture, Figure~\ref{Fig1}(a), which allows to selectively dope each BLG with electrons or holes (see details of sample preparation in the Supporting Information Section S1); the measurements were performed at $4.2\,\mbox{K}$. Tunneling in the zero-bias limit $V_\mathrm{b}\rightarrow0$ in these devices was studied in our previous papers \cite{Vdovin2024,Sokolik2025}. As shown in the maps of zero-bias tunneling conductance $G=(dI/dV_\mathrm{b})|_{V_\mathrm{b}=0}$ in Figure~\ref{Fig1}(c,d), tuning the gate voltages brings the system to the regime of allowed ($G>0$) or forbidden ($G=0$) tunneling, depending on whether the mutually shifted electron dispersions of top and bottom BLGs intersect or not at the common Fermi level. Since the lower gates of two devices have different designs, namely 290-nm $\mathrm{SiO}_2$ on top of Si for Device 1 and 50-nm hexagonal boron nitride (hBN) layer on top of graphite for Device 2, the values of $V_\mathrm{bg}$ needed to achieve the same doping level of bottom BLG in Device 2 is about 5 times larger that in Device 1. 

Application of a bias voltage $V_\mathrm{b}$ moves the Fermi levels of two BLGs apart and involves those parts of electron dispersions which are constrained between the Fermi levels, i.e. electron states which are occupied in one BLG and empty in the other BLG. Electron states involved in the tunneling are located in vicinity (due to nonzero energy broadening) of the lines in energy-momentum space where the dispersions intersect each other between the Fermi levels. As exemplified in Figure~\ref{Fig1}(b), such intersections can involve like bands (both conduction or both valence bands), giving rise to the like-band tunneling, of unlike bands (conduction and valence bands in different BLGs), which open the channel of unlike-band tunneling. The presence of these two tunneling channels is a key to understand the I-V characteristics $I(V_\mathrm{b})$, as will be shown below.

Our experimental I-V characteristics are analyzed using the theoretical model, which takes into account electrostatics of the dual-gated system, quantum capacitance effects, energy level broadening, finite temperature, self-consistent gap opening in each BLG, and the tunneling matrix elements which depend on a distribution of electron wave function over graphene sublayers of each BLG. The theoretical model is described in detail in the Supporting Information Section S2.

\begin{figure*}[t]
\centering
\includegraphics[width=0.9\textwidth]{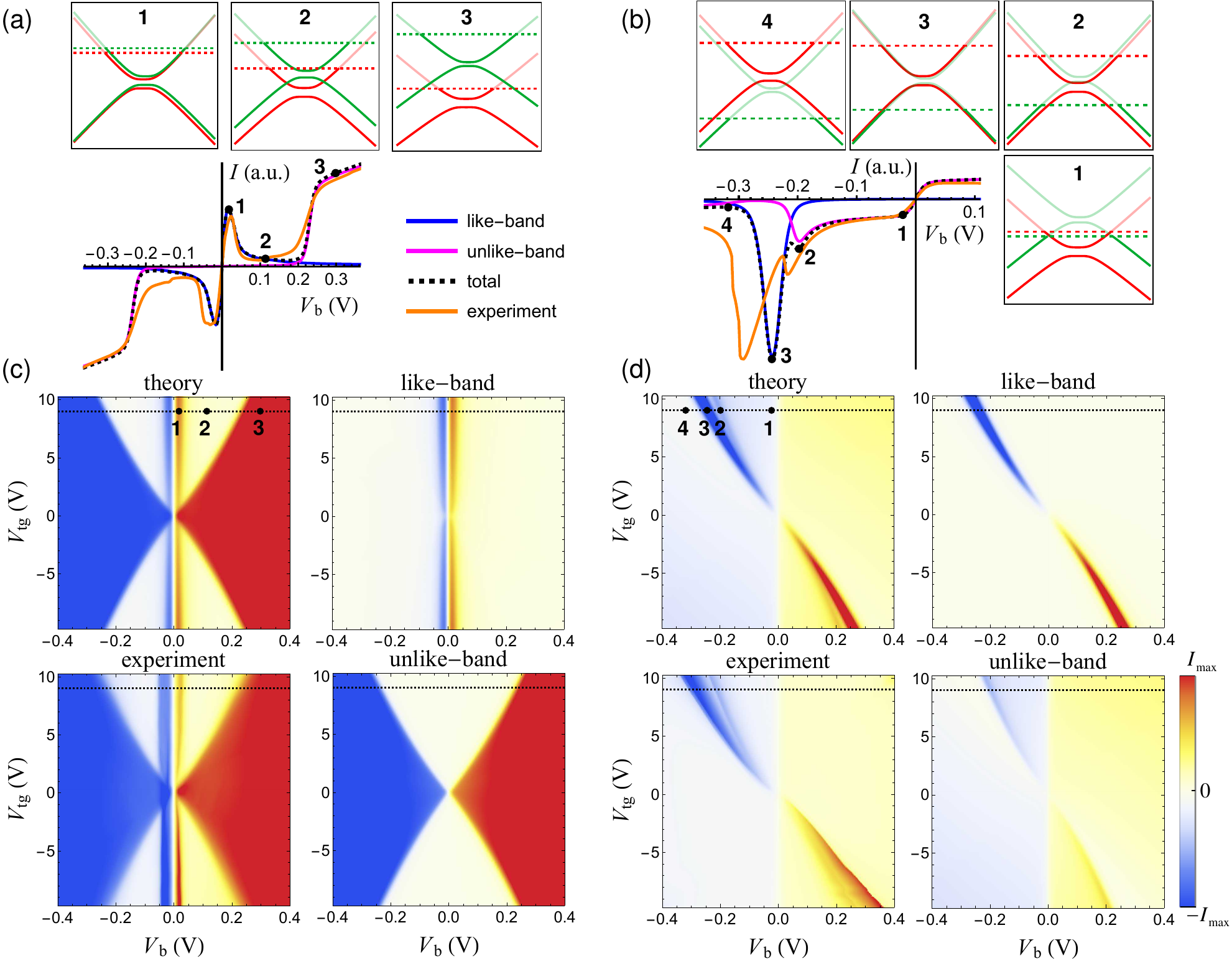}
\caption{Tunneling in Device 1. (a) I-V characteristic at $V_\mathrm{tg}=9\,\mbox{V}$, $V_\mathrm{bg}=55.2\,\mbox{V}$ where carrier densities in both BLGs are equal at zero bias. (b) I-V characteristic at $V_\mathrm{tg}=9\,\mbox{V}$, $V_\mathrm{bg}=-54.8\,\mbox{V}$ where carrier densities are opposite. Insets in (a,b) shows calculated electron dispersions in top (green) and bottom (red) BLG at selected points. Dotted lines mark the Fermi level locations, and the parts of dispersions corresponding to unoccupied electron states are semitransparent. (c) Two-dimensional maps of measured and calculated tunneling current as functions of $V_\mathrm{b}$ and $V_\mathrm{tg}$ (left panels), and calculated current divided into like- and unlike-band components (right panels); $V_\mathrm{bg}$ changes with $V_\mathrm{tg}$ to satisfy the condition of equal densities in both BLGs at $V_\mathrm{b}=0$. (d) The same as (c) but when $V_\mathrm{tg}$ and $V_\mathrm{bg}$ are changed to satisfy the condition of opposite densities. Dashed lines in (c,d) indicate the cuts where  the I-V characteristics in (a,b) are drawn.}
\label{Fig2}
\end{figure*}

Figure~\ref{Fig2} shows the measured and calculated I-V characteristics for Device 1 with the twist angle $\theta=0.13^\circ$. Similar structures with $\theta\approx0^\circ$ were previously studied in Refs.~\cite{Fallahazad2015,Kim2016,Burg2017}. The left (a,c) and right (b,d) parts of the figure present two qualitatively different types of I-V characteristics. The first type is a symmetric characteristic, Figure~\ref{Fig1}(a), where $I$ is an odd function of $V_\mathrm{b}$. It emerges at those combinations of gate voltages $V_\mathrm{tg}$, $V_\mathrm{bg}$ where carrier densities in both BLGs coincide at zero bias (red line in Figure~\ref{Fig1}(c)). Initial increase of $V_\mathrm{b}$ pushes the Fermi levels apart opening the way to electron tunneling between the like bands with almost aligned dispersions, so the resonant tunneling peak is formed (point 1). Further increase of $V_\mathrm{b}$ misaligns the dispersions along energy axis, thus the tunneling is suppressed (point 2). At even higher $V_\mathrm{b}$, the overlap of unlike bands falls within an interval between the Fermi levels, and the unlike-band tunneling channel opens giving rise to a step in the I-V characteristic (point 3). Cross-sections of calculated electron dispersions along the momentum displacement vector separating the Dirac points of two BLGs, shown in the insets of Figure~\ref{Fig1}(a), clearly demonstrate the formation mechanism of the observed features of I-V characteristic.

Figure 2(c) shows the two-dimensional theoretical and experimental current maps as a function of the bias voltage $V_\mathrm{b}$ and the gate voltages $V_\mathrm{tg}$, $V_\mathrm{bg}$ changing simultaneously to satisfy the condition of equal carrier densities in both BLGs at $V_\mathrm{b}=0$ (along red line in Figure~\ref{Fig1}(c)). Each horizontal cut at $V_\mathrm{tg}=\mathrm{const}$ behaves similarly to Figure~\ref{Fig2}(a): the ambipolar peaks at small $|V_\mathrm{b}|$ (due to resonant like-band current) and steps (in unlike-band current) at larger threshold voltages  |$V_\mathrm{b}|$, which increase with the doping level.

The second type of I-V characteristics is asymmetric, and occurs when the doping levels at $V_\mathrm{b}=0$ significantly differ. As extreme example, we show in Figure 2(b) the case of opposite charge carrier densities in two BLGs. At $V_\mathrm{b}>0$, tunneling between unlike bands occurs immediately, since they already intersect at $V_\mathrm{b}=0$ at the common Fermi level. At $V_\mathrm{b}<0$ the situation is more complicated. After opening of the unlike-band tunneling (point 1), further increase of $|V_\mathrm{b}|$ aligns the van Hove singularities of the density of states residing at the band extrema  \cite{Kim2013,Joucken2021} of the opposite bands. This gives rise to the peak of unlike-band current (point 2). Then the dispersions in both BLGs align in energy, which suppresses the unlike-band tunneling, but allows for the resonant like-band tunneling (point 3). Note that separate sub-peaks of unlike- and like-band currents combine to a split peak of the total current, in agreement with experiment. At even higher $|V_\mathrm{b}|$, the unlike-band overlap reappears (point 4). The two-dimensional maps in Figure~\ref{Fig2}(d) plotted versus $V_\mathrm{b}$ and gate voltages $V_\mathrm{tg}$, $V_\mathrm{bg}$ changes simultaneously to satisfy the condition of equal density but opposite signs of the carriers in both BLGs  (blue line in Figure~\ref{Fig1}(c)) demonstrate the same features: the step of unlike-band current at $V_\mathrm{b}\approx0$, the peak of unlike-band current (whose distance from the origin increases with the doping) followed by more intense peak of like-band current. Additional tunneling current maps and I-V characteristics are presented in the Supporting Information Sections S3 and S4.

\begin{figure}[t]
\centering
\includegraphics[width=0.5\textwidth]{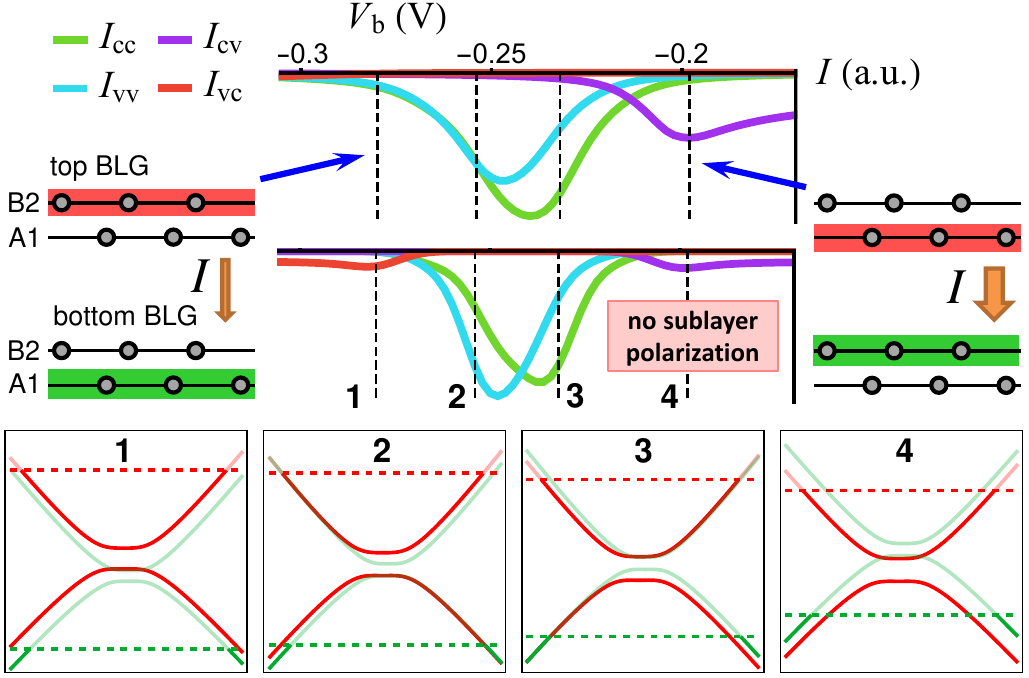}
\caption{Magnified part of I-V calculated characteristic of Device 1 at $V_\mathrm{tg}=9\,\mbox{V}$, $V_\mathrm{bg}=-54.8\,\mbox{V}$ where carrier densities in both BLGs are opposite at $V_\mathrm{b}=0$. The current is decomposed over initial and final electrons bands into the like-band $I_\mathrm{cc}$, $I_\mathrm{vv}$ and unlike-band $I_\mathrm{cv}$, $I_\mathrm{vc}$ components. The lower panel present calculations without taking into account the sublayer polarization of electron wave functions. The bottom insets 1-4 show alignment of electron dispersions at those bias voltages (dashed lines on the plot) where the like- or unlike-band overlaps of the van Hove singularities occur. The schemes on the sides show wave function distributions over BLG sublattices in points 1 and 4.}
\label{Fig3}
\end{figure}

\begin{figure*}[t]
\centering
\includegraphics[width=0.9\textwidth]{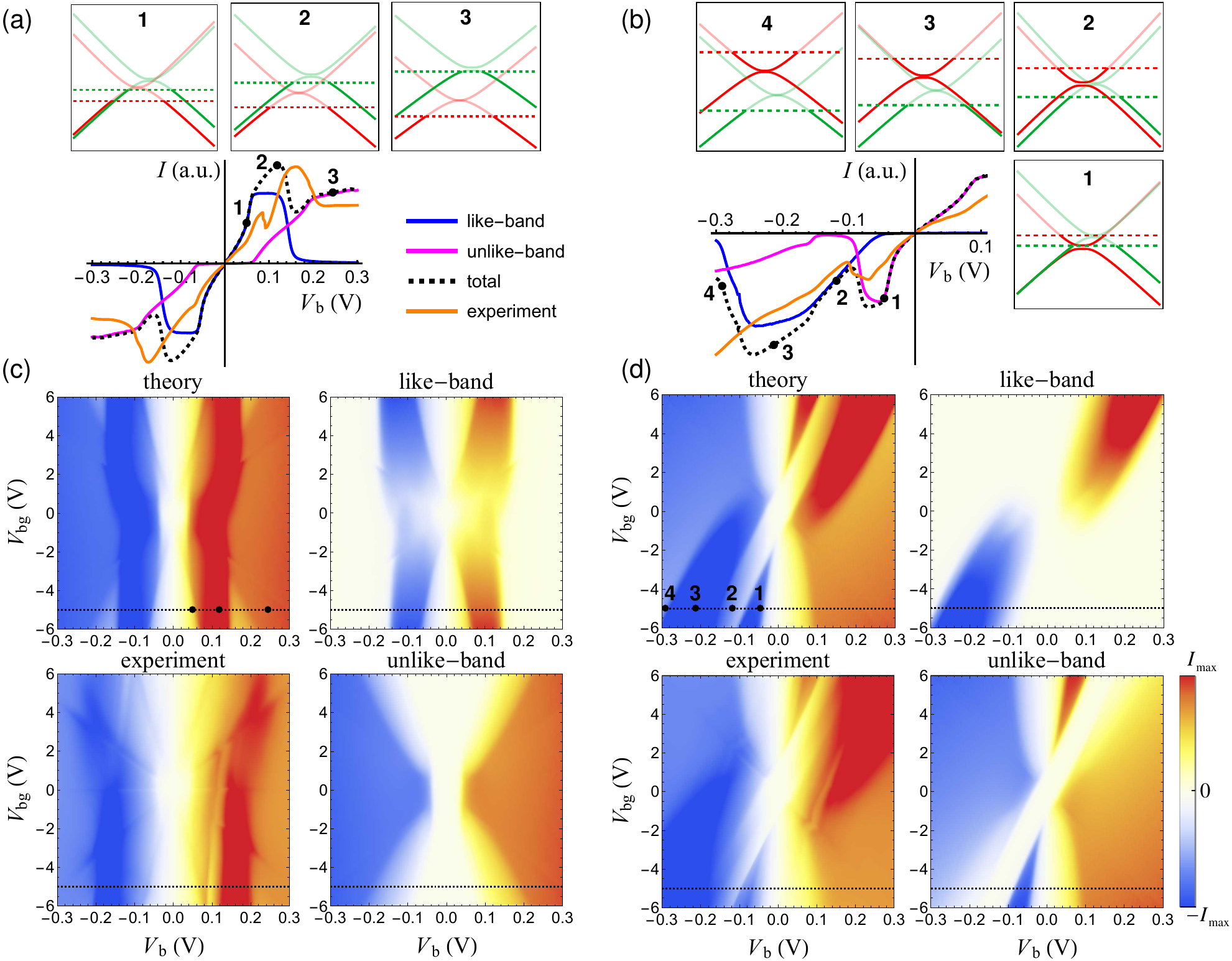}
\caption{Tunneling in Device 2. (a) I-V characteristic at $V_\mathrm{tg}=-3.83\,\mbox{V}$, $V_\mathrm{bg}=-5\,\mbox{V}$ where carrier densities in both BLGs are equal at zero bias. (b) I-V characteristic at $V_\mathrm{tg}=4.39\,\mbox{V}$, $V_\mathrm{bg}=-5\,\mbox{V}$ where carrier densities are opposite. Insets in (a,b) shows calculated electron dispersions in top (green) and bottom (red) BLG at selected points. Dotted lines mark the Fermi level locations, and the parts of dispersions corresponding to unoccupied electron states are semitransparent. (c) Two-dimensional maps of the measured and calculated tunneling current as functions of $V_\mathrm{b}$ and $V_\mathrm{bg}$ (left panels), and calculated current divided into like- and unlike-band components (right panels); $V_\mathrm{tg}$ changes with $V_\mathrm{bg}$ to satisfy the condition of equal densities in both BLGs. (d) The same as (c) but when $V_\mathrm{tg}$ and $V_\mathrm{bg}$ are changed to satisfy the condition of opposite densities. Dashed lines in (c,d) indicate the cuts where the I-V characteristics in (a,b) are drawn.}
\label{Fig4}
\end{figure*}

Figure~\ref{Fig3} presents detailed picture of calculated peaks in the asymmetric I-V characteristic, revealing the  role of van Hove singularities of electron density of states at band extrema of each BLG. The calculated current is resolved here by energy bands of the tunneling electrons: $I_\mathrm{cc}$ ($I_\mathrm{vv}$) is the like-band current between the pair of conduction (valence) bands, and $I_\mathrm{cv}$ ($I_\mathrm{vc}$) is the unlike-band current between conduction (valence) band of top BLG and valence (conduction) band of bottom BLG. The resonant peaks in $I_\mathrm{cc}$ and $I_\mathrm{vv}$ arise when dispersions of like bands are aligned (points 2 and 3). The overlap of van Hove singularities of the unlike bands (points 1 and 4) should seemingly provide two symmetric peaks of the unlike-band currents, but actually the picture is strongly asymmetric: only in point 4 the peak of $I_\mathrm{cv}$ develops, but in the complementary point 1 the corresponding current $I_\mathrm{vc}$ vanishes.

The origin of such asymmetry lies in the sublayer polarization of the electron wave functions \cite{Kim2013,Joucken2021}, because calculations performed without taking it into account (lower plot in Figure~\ref{Fig3}) demonstrate the symmetric picture of two peaks both in $I_\mathrm{cv}$ and $I_\mathrm{vc}$ (see similar calculations in Ref.~\cite{delaBarrera2015}). The sublayer polarization greatly enhances the peak of $I_\mathrm{cv}$ by drawing electron wave functions from different BLG closer to each other and completely suppresses the peak of  $I_\mathrm{vc}$ by pulling the wave functions apart. This effect arising at nonzero bias in Device 1 complements the asymmetric resonant peaks observed at zero bias in Device 2 \cite{Vdovin2024,Sokolik2025} and having the same origin.

Device 2 with the multiply increased twist angle $\theta=0.73^\circ$ demonstrates the qualitatively similar I-V characteristics, although with wider peaks, as shown in Figure~\ref{Fig4}. Again, when the carrier densities coincide at zero bias (equal-density diagonal in Figure~\ref{Fig1}(d)), the I-V characteristic is symmetric, as shown in Figure~\ref{Fig4}(a). Increasing $V_\mathrm{b}$ opens a tunneling channel between the like bands (point 1), although not immediately (in contrast to Device 1) due to momentum mismatch, which should be compensated by a shift of dispersions along the energy axis. Further increase of $V_\mathrm{b}$ opens the unlike-band tunneling in the interval between the Fermi levels (point 2). At even higher $V_\mathrm{b}$, the overlap of unlike bands enlarges, as does the corresponding current component, but the like-band current disappears because the shift of dispersions along energy axis becomes too strong (point 3). The peak of the like-band current and the step of unlike-band one is much smoother than in Device 1. The two-dimensional theoretical and experimental maps of the current in Figure~\ref{Fig4}(c) as functions of $V_\mathrm{b}$ and $V_\mathrm{bg}$ along the equal-density diagonal demonstrate the symmetric butterfly-like picture. The maps of like- and unlike-band currents are similar to those in Device 1 (Figure~\ref{Fig2}), but the allowed ||-shaped regions of the like-band tunneling are wider, and the X-shaped thresholds for the unlike-band tunneling are smoother.

The asymmetric I-V characteristic takes place when the doping levels significantly differ at zero bias, as shown in Figure~\ref{Fig4}(b) for a point at the opposite-density diagonal. It is qualitatively similar to those in Device 1 (Figure~\ref{Fig2}(b)), although the unlike-band current reaches its peak value (point 1) much earlier than the like-band current (point 3), and both peaks are wider. Note that alignment of electron dispersions in energy (point 2) does not lead to current enhancement, in contrast to Device 1, because of significant momentum mismatch between the dispersions. This mismatch is the origin of enhanced separation of the like- and unlike-band current peaks (points 1 and 3), which leads to markedly split peak of the total current, in contrast to Device 1, where the splitting is barely seen. Besides relatively sharp peak (point 1), the unlike-band current demonstrates very smooth rise at large $|V_\mathrm{b}|$ (point 4), whose delay can be attributed to sublayer polarized van Hove singularities \cite{Vdovin2024,Sokolik2025}. The two-dimensional maps along the opposite-density diagonal in Figure~\ref{Fig4}(d) are shaped mainly by two forbidden regions for the unlike-band tunneling: one is close to the vertical axis (where too narrow interval between the Fermi energies does not enclose the interband overlap of dispersions), and the other extends along the /-shaped line (when the bands align in energy, so the interband overlap is absent). Our calculations generally agree with the experiment. The maps of differential tunneling conductance $dI/dV_\mathrm{b}$ for Device 2 along the opposite- and equal-density diagonals are shown in the abstract, and additional maps and I-V characteristics are presented in the Supporting Information Sections S3 and S4.

In summary, we have analyzed theoretically and experimentally the tunneling characteristics between closely spaced BLG sheets mutually twisted on different small angles. Combination of momentum (due to the twist) and energy (due to the bias voltage) relative displacements of electron dispersions in two BLGs gives rise to diverse types of I-V characteristics depending on the gate voltages and on the twist angle. Our analysis reveals the existence of two kinds of characteristics: the symmetric one when both BLGs are doped equally at zero bias, and asymmetric one when the doping levels at zero bias significantly differ. The like-band tunneling typically demonstrates resonant tunneling peaks when well aligned parts of electron dispersions in two BLGs fall between the Fermi levels. The unlike-band tunneling typically gives rise to thresholds in I-V characteristics when the interband intersection of dispersions appears and enters an interval between the Fermi levels. In asymmetric I-V characteristics, the resonant peak of this current arises when van Hove singularities from unlike bands overlap, however the complementary peak with the reversed band order is suppressed by sublayer polarization of electron wave functions. At vanishing twist angle, separation between the like- and unlike-band peaks in asymmetric I-V characteristics is approximately equal to band gap induced in both BLGs by vertical electric field. Both kinds of resonant current peaks give rise to negative differential resistance, which is of practical interest.

In contrast to the existing literature \cite{Britnell2013,Mishchenko2014,Zhang2023,Zhang2025,Zhang2026,Prasad2021,Kuzmina2021}, our analysis of tunneling characteristics includes the sample with larger twist angle introducing a sizable momentum displacement between electron dispersion. The qualitative features of I-V characteristics discussed above persist at larger twist angles, although resonant peaks of like-band current and forbidden regions for unlike-band current become wider. The gradual progression of the tunneling current maps with increasing twist angle is shown in the Supporting Information Section S5. Our theoretical model reproduces the measurements with reasonable accuracy and provides insight to the origin of observed features. Previous theoretical studies of I-V characteristics of bilayer graphene heterostructures \cite{Fallahazad2015,Kim2016,Burg2017,Zhang2025,Prasad2021,delaBarrera2015,Zhang2025,Burg2018} did not take into account the gap opening and sublayer polarization. The systematic analysis of tunneling current in different conditions and disclosure of its formation mechanisms may prove useful for finding the most pronounced regions of negative differential resistance and for studying the limits of tunneling current control using gates, gap opening, and wave function redistribution.

\section*{Acknowledgements}

The experimental work of E.E.V, Y.N.K, S.V.M was supported by the Russian Science Foundation (Grant No. 23-12-00115-P). A.A.S. acknowledges receiving support for his work on the theoretical model from the Program of Basic Research of the Higher School of Economics. A.F.A. acknowledges the Basis Foundation for the support of numerical calculations. M.A.K. acknowledges receiving support for his work on sample preparation from the Russian Science Foundation (Grant No. 21-79-20225-P) and the internal funding program of the Center for Neurophysics and Neuromorphic Technologies.

\section*{Supporting information}

The Supporting Information is available free of charge at...

Description of sample preparation, description of the theoretical model, additional calculated and measured 2D maps of tunneling current and differential conductance, additional I-V characteristics, demonstration of the twist angle progression (PDF).

\printbibliography

\end{document}


\maketitle

\section{Preparation of the samples}

\begin{figure}[!b]         
\centering
\includegraphics[width=\textwidth]{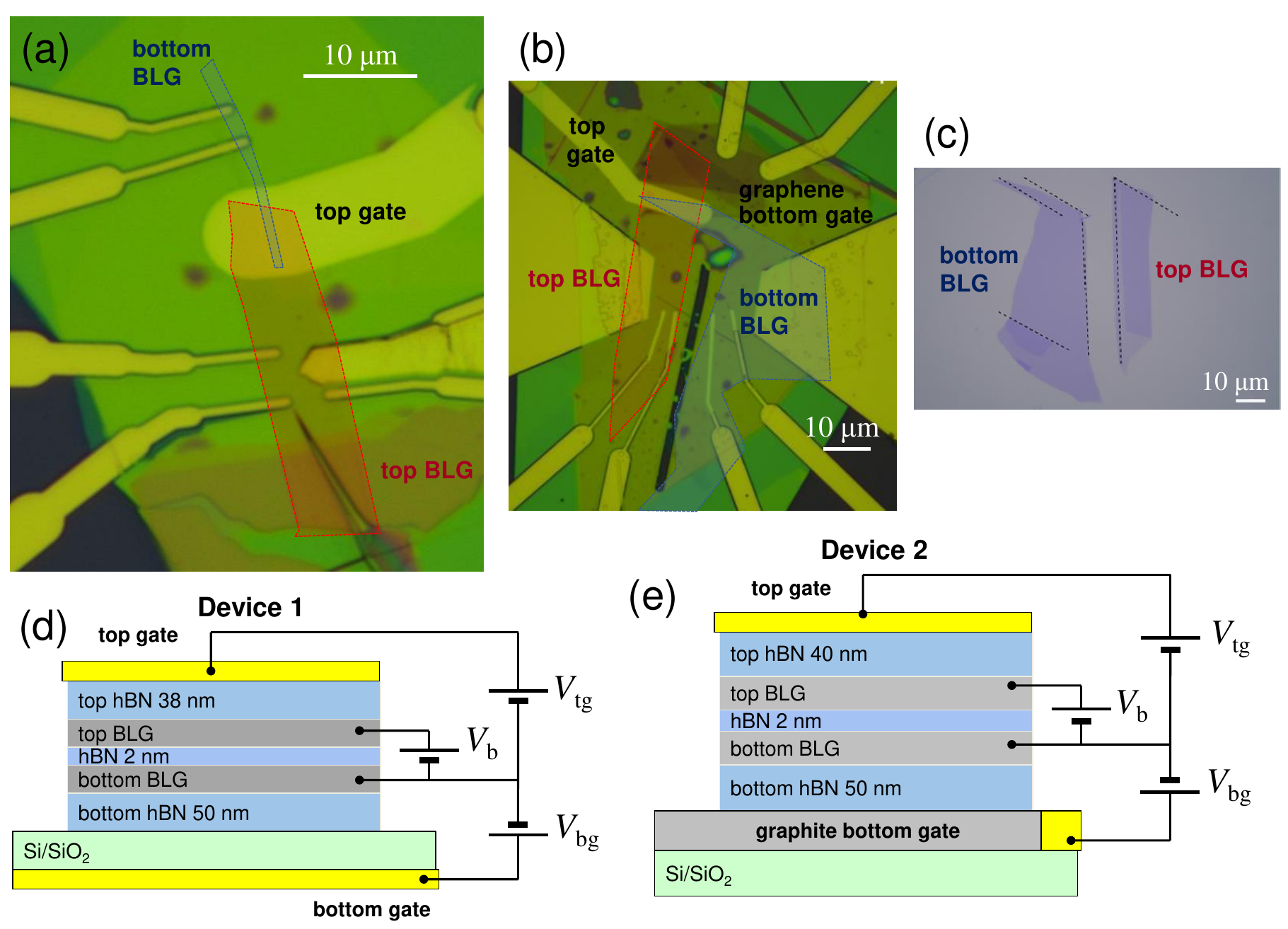}
\caption{Optical micrographs of BLG-hBN-BLG (a) Device 1 and (b) Device 2. Red and blue dashed contours indicate  the top and bottom BLG flakes, respectively.
(c) Optical microscope image of mechanically exfoliated BLG flakes of Device 2 before its assembling. Dashed lines mark the edges of BLG flakes. (d,e) Cross-sectional schematics of the final Device 1 and Device 2, respectively, and the electric biasing scheme used for the two-point interlayer tunneling current measurements.}
\label{FigS1}
\end{figure}

In this paper, we investigated two devices. Photographs of the devices are shown in Figure~\ref{FigS1}: Device 1 (a) and Device 2 (b). The tunneling devices were fabricated using a dry transfer method applied to micromechanically exfoliated layers of Bernal bilayer graphene (BLG) and hexagonal boron nitride (hBN) of varying thicknesses from bulk graphite and hBN crystals, respectively. The process began with mechanical exfoliation of bilayer Bernal graphene and hBN flakes on a $\mathrm{Si/SiO}_2$ substrate. The final van der Waals heterostructure was assembled layer by layer (from top to bottom) using a PC/PDMS stamp and finally transferred onto the target substrate. To ensure a close-to-zero twist angle between the top and bottom BLG, adjacent bilayer graphene flakes (see Figure 1(c) for Device 2) obtained from a single ruptured flake were used. 

The final structure for both samples consisted of a bottom BLG electrode, an hBN tunnel barrier, and a top BLG electrode. For clarity, the bottom (blue) and top (red) BLG flakes are shown in Figures~\ref{FigS1}(a) and (b) using dashed lines and translucent shading. The bottom gate differs in the two heterostructures: in Device 1 (Figure~\ref{FigS1}(d)), a silicon substrate separated from the BLG electrode by a 290 nm-thick $\mathrm{SiO}_2$ dielectric layer served as the gate, while in Device 2 (Figure~\ref{FigS1}(e)), a graphite flake separated from the BLG by a 48 nm-thick hBN layer served as the gate. The twist angle in the heterostructures in Device 1 was approximately $0.7^\circ$, while in Device 2 it was approximately $0.1^\circ$. In Device 2, the bottom BLG layer was rotated by approximately $1^\circ$ relative to the bottommost hBN layer, and the top BLG layer was rotated by approximately $3^\circ$ relative to the topmost hBN layer. Next, Cr/Au edge contacts were created on the bottom and top BLG layers using electron beam lithography, followed by hBN etching, metal deposition, and a lift-off process. $\mathrm{SF}_6$ reactive ion etching was used to selectively etch the top hBN layer before metal deposition. The top hBN encapsulation layer was further coated with a Cr/Au pad, which served as the top gate electrode. The active areas of the devices (overlap of the top and bottom BLG) were approximately $1.4\,\mu\mbox{m}^2$ for Device 1 and $18\,\mu\mbox{m}^2$ for Device 2. 

\section{Theoretical model}
\subsection{Electrostatics}
\begin{figure}[t]         
\centering
\includegraphics[width=0.6\textwidth]{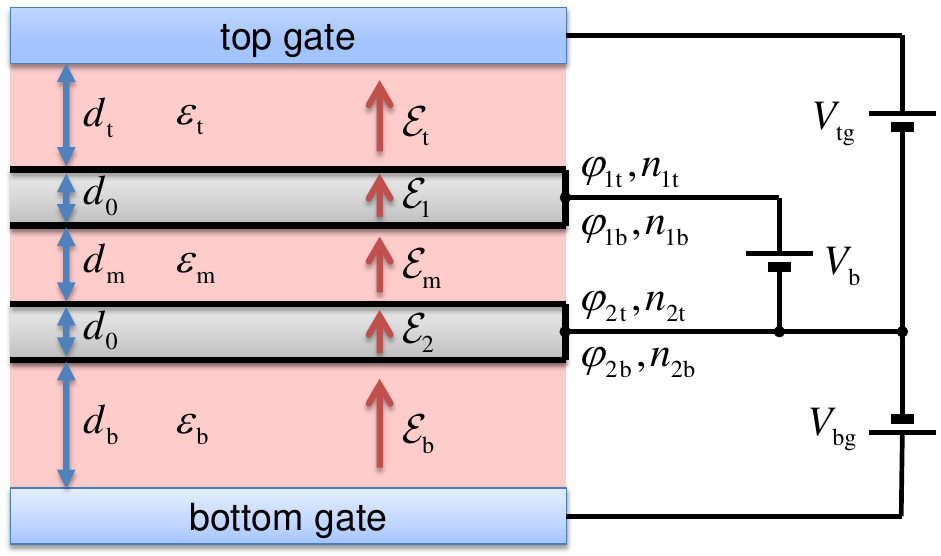}
\caption{Electrostatic schematic of the dual-gated system. Top and bottom BLGs are resolved over graphene sublayers;  the distances $d_i$ and electric fields $\mathcal{E}_i$ between these sublayers and gates are indicated. The dielectric constants of three dielectric spacers are $\varepsilon_i$ (we assume $\varepsilon=1$ inside each BLG). Each sublayer is also characterized by its electrostatic potential $\varphi_i$ and electron density $n_i$.}
\label{FigS2}
\end{figure}

The model used to calculate the tunneling current is similar to that used in our previous paper \cite{Sokolik2025}. Consider the system of two bilayer graphene (BLG) sheets, each resolved by graphene sublayers (Fig.~\ref{FigS2}). The Gauss theorem for each sublayer read
\begin{align}
-\varepsilon_\mathrm{t}\mathcal{E}_\mathrm{t}+\mathcal{E}_1=4\pi en_\mathrm{1t},\qquad
-\mathcal{E}_1+\varepsilon_\mathrm{m}\mathcal{E}_\mathrm{m}=4\pi en_\mathrm{1b},\label{el1}\\
-\varepsilon_\mathrm{m}\mathcal{E}_\mathrm{m}+\mathcal{E}_2=4\pi en_\mathrm{2t},\qquad
-\mathcal{E}_2+\varepsilon_\mathrm{b}\mathcal{E}_\mathrm{b}=4\pi en_\mathrm{2b},\label{el2}
\end{align}
where $n_{i\mathrm{t}}$ and $n_{i\mathrm{b}}$ are electron densities at top and bottom sublayers of top ($i=1$) and bottom ($i=2$) BLG. The electric fields $\mathcal{E}_i$ and dielectric constants $\varepsilon_i$ are shown in Fig.~\ref{FigS2}. The electron charge is taken as $-e$, where $e>0$. Electrons in top ($i=1$) and bottom ($i=2$) BLG have electrochemical potentials $-e(\varphi_{i\mathrm{t}}+\varphi_{i\mathrm{b}})/2+\mu_i$ consisting of mean electrostatic energy of its sublayers and chemical potential $\mu_i$ counted from its Dirac point. The bias voltage equals to their difference:
\begin{equation}
e\frac{\varphi_{1\mathrm{t}}+\varphi_{1\mathrm{b}}}2-\mu_1-e\frac{\varphi_{2\mathrm{t}}+\varphi_{2\mathrm{b}}}2+\mu_2=eV_\mathrm{b}.\label{el3}
\end{equation}

Similarly, for differences of electrochemical potentials in top gate and top BLG, and in bottom gate and bottom BLG, we obtain
\begin{equation}
e\varphi_\mathrm{tg}-e\frac{\varphi_{1\mathrm{t}}+\varphi_{1\mathrm{b}}}2+\mu_1=e(V_\mathrm{tg}-V_\mathrm{b}),\qquad
e\varphi_\mathrm{bg}-e\frac{\varphi_{2\mathrm{t}}+\varphi_{2\mathrm{b}}}2+\mu_2=eV_\mathrm{bg},\label{el4}
\end{equation}
where $\varphi_\mathrm{tg}$ and $\varphi_\mathrm{bg}$ are electrostatic potentials of the top and bottom gates. We neglect electron chemical potentials in the metallic gates (i.e. consider them as constants) assuming their overwhelmingly high densities of states. We can also relate potential differences to electric fields,
\begin{equation}
\varphi_{1\mathrm{t}}-\varphi_\mathrm{tg}=d_\mathrm{t}\mathcal{E}_\mathrm{t},\qquad\varphi_{1\mathrm{b}}-\varphi_{1\mathrm{t}}=d_0\mathcal{E}_1,\qquad\varphi_{2\mathrm{t}}-\varphi_{1\mathrm{b}}=d_\mathrm{m}\mathcal{E}_\mathrm{m},\label{el5}\\
\end{equation}
\begin{equation}
\varphi_{2\mathrm{b}}-\varphi_{2\mathrm{t}}=d_0\mathcal{E}_2,\qquad\varphi_\mathrm{bg}-\varphi_{2\mathrm{b}}=d_\mathrm{b}\mathcal{E}_\mathrm{b},\label{el6}
\end{equation}
and take into account that electron energy difference $U_i$ between upper and lower sublayers of the $i$-th BLG is related to the difference of their potentials as
\begin{equation}
-e(\varphi_{1\mathrm{t}}-\varphi_{1\mathrm{b}})=U_1,\qquad
-e(\varphi_{2\mathrm{t}}-\varphi_{2\mathrm{b}})=U_2.\label{el7}
\end{equation}

Excluding electric fields $\mathcal{E}_i$ from Eqs.~(\ref{el1})-(\ref{el2}), (\ref{el5})-(\ref{el6}), and introducing the capacitances per unit area
\begin{equation}
C_\mathrm{t}=\frac{\varepsilon_\mathrm{t}}{4\pi ed_\mathrm{t}},\qquad C_\mathrm{m}=\frac{\varepsilon_\mathrm{m}}{4\pi ed_\mathrm{m}},\qquad C_\mathrm{b}=\frac{\varepsilon_\mathrm{b}}{4\pi ed_\mathrm{b}},\qquad C_0=\frac1{4\pi ed_0},\label{el8}
\end{equation}
we obtain
\begin{align}
C_\mathrm{t}(\varphi_\mathrm{tg}-\varphi_{1\mathrm{t}})+C_0(\varphi_{1\mathrm{b}}-\varphi_{1\mathrm{t}})=en_{1\mathrm{t}},\qquad
C_0(\varphi_{1\mathrm{t}}-\varphi_{1\mathrm{b}})+C_\mathrm{m}(\varphi_{2\mathrm{t}}-\varphi_{1\mathrm{b}})=en_{1\mathrm{b}},\label{el9}\\
C_\mathrm{m}(\varphi_{1\mathrm{b}}-\varphi_{2\mathrm{t}})+C_0(\varphi_{2\mathrm{b}}-\varphi_{2\mathrm{t}})=en_{2\mathrm{t}},\qquad
C_0(\varphi_{2\mathrm{t}}-\varphi_{2\mathrm{b}})+C_\mathrm{b}(\varphi_\mathrm{bg}-\varphi_{2\mathrm{b}})=en_{2\mathrm{b}}.\label{el10}
\end{align}
Then, excluding the potentials $\varphi_i$ from Eqs.~(\ref{el3})-(\ref{el4}), (\ref{el7}), (\ref{el9})-(\ref{el10}), we obtain
\begin{align}
\left(C_0+\frac12C_\mathrm{t}\right)\frac{U_1}e-C_\mathrm{t}\frac{\mu_1}e+C_\mathrm{t}(V_\mathrm{tg}-V_\mathrm{b})=en_{1\mathrm{t}},\\
-\left(C_0+\frac12C_\mathrm{m}\right)\frac{U_1}e-C_\mathrm{m}\frac{U_2}{2e}-C_\mathrm{m}\frac{\mu_1}e+C_\mathrm{m}\frac{\mu_2}e-C_\mathrm{m}V_\mathrm{b}=en_{1\mathrm{b}},\\
C_\mathrm{m}\frac{U_1}{2e}+\left(C_0+\frac12C_\mathrm{m}\right)\frac{U_2}e+C_\mathrm{m}\frac{\mu_1}e-C_\mathrm{m}\frac{\mu_2}e+C_\mathrm{m}V_\mathrm{b}=en_{2\mathrm{t}},\\
-\left(C_0+\frac12C_\mathrm{b}\right)\frac{U_2}e-C_\mathrm{b}\frac{\mu_2}e+C_\mathrm{b}V_\mathrm{bg}=en_{2\mathrm{b}}.
\end{align}
Taking sums and differences of these equations, and introducing the total electron densities at top and bottom BLGs, as well as sublayer density imbalances,
\begin{equation}
n_1=n_{1\mathrm{t}}+n_{1\mathrm{b}},\qquad
n_2=n_{2\mathrm{t}}+n_{2\mathrm{b}},\qquad
\Delta n_1=n_{1\mathrm{t}}-n_{1\mathrm{b}},\qquad
\Delta n_2=n_{2\mathrm{t}}-n_{2\mathrm{b}},
\end{equation}
we obtain
\begin{align}
\left(C_\mathrm{t}-C_\mathrm{m}\right)\frac{U_1}{2e}-C_\mathrm{m}\frac{U_2}{2e}-(C_\mathrm{t}+C_\mathrm{m})\frac{\mu_1}e+C_\mathrm{m}\frac{\mu_2}e+C_\mathrm{t}V_\mathrm{tg}-(C_\mathrm{m}+C_\mathrm{t})V_\mathrm{b}=en_1,\label{el16}\\
C_\mathrm{m}\frac{U_1}{2e}+\left(C_\mathrm{m}-C_\mathrm{b}\right)\frac{U_2}{2e}+C_\mathrm{m}\frac{\mu_1}e-(C_\mathrm{m}+C_\mathrm{b})\frac{\mu_2}e+C_\mathrm{b}V_\mathrm{bg}+C_\mathrm{m}V_\mathrm{b}=en_2,\label{el17}
\end{align}
\begin{align}
\left(2C_0+\frac12C_\mathrm{t}+\frac12C_\mathrm{m}\right)\frac{U_1}e&+C_\mathrm{m}\frac{U_2}{2e}\nonumber\\
&+(C_\mathrm{m}-C_\mathrm{t})\frac{\mu_1}e-C_\mathrm{m}\frac{\mu_2}e+C_\mathrm{t}V_\mathrm{tg}+(C_\mathrm{m}-C_\mathrm{t})V_\mathrm{b}=e\Delta n_1,\label{el18}
\end{align}
\begin{align}
C_\mathrm{m}\frac{U_1}{2e}+\left(2C_0+\frac12C_\mathrm{m}+\frac12C_\mathrm{b}\right)&\frac{U_2}e\nonumber\\
&+C_\mathrm{m}\frac{\mu_1}e-(C_\mathrm{m}-C_\mathrm{b})\frac{\mu_2}e-C_\mathrm{b}V_\mathrm{bg}+C_\mathrm{m}V_\mathrm{b}=e\Delta n_2.\label{el19}
\end{align}
To close this system of equations, we need constitutive relations connecting electron densities $n_{1,2}$ and imbalances $\Delta n_{1,2}$ to chemical potentials $\mu_{1,2}$ and field-induced gaps $U_{1,2}$. These relations are presented in the next subsection.

\subsection{Electron properties of BLG}

The tight binding model of BLG \cite{McCann2013} reproduces a number of important features of electron dispersion: 4 energy bands, quadratic dispersion of low-energy bands turning into a linear one at large momenta, electron-hole asymmetry, trigonal warping, and gap opening in a vertical electric field. We use the simplified model, which is based on analytically tractable $(2\times2)$ Hamiltonian for the lowest-energy bands and takes into account the dispersion hyperbolicity, the electron-hole asymmetry, and the gap opening.

The dispersion laws
\begin{equation}
E_\pm^0=\mp\frac{\gamma_1}2\pm\sqrt{(v\pm v_4)^2\hbar^2k^2+\left(\frac{\gamma_1}2\right)^2}
\end{equation}
interpolate between quadratic behavior $E_\pm^0\approx\pm\hbar^2k^2/2m_\pm$ at low momenta with the effective masses $m_\pm=\gamma_1/2(v\pm v_4)^2$ and linear asymptotic $E_\pm^0\approx\pm(v\pm v_4)\hbar k$ at high momenta. Here $\gamma_1$ is the vertical hopping integral, $v$ is the graphene Fermi velocity, and $v_4$ is its electron-hole asymmetry. The effective mass at low momenta and the Fermi velocity at high momenta are both asymmetric in this model, which agrees with experiments and full tight binding model \cite{Zou2011}. To take into account the gap opening, we use the following approximate Hamiltonian:
\begin{equation}
H=\left(\begin{array}{cc}
\displaystyle\frac{E_+^0+E_-^0-U}2&\displaystyle\frac{-E_+^0+E_-^0}2e^{-2i\xi\phi}\\\\
\displaystyle\frac{-E_+^0+E_-^0}2e^{2i\xi\phi}&\displaystyle\frac{E_+^0+E_-^0+U}2
\end{array}\right).\label{S_H}
\end{equation}
Two rows and columns correspond to sublattices A1 and B2 composing the low-energy electron states of BLG, $U$ is the difference of electron potential energies between the upper (B2) and lower (A1) sublayers, $\xi=\pm1$ for valleys $\mathbf{K}$ and $\mathbf{K}'$, and $\phi$ is the azimuthal angle of electron momentum $\mathbf{k}$. At low momenta $k\rightarrow0$, the Hamiltonian (\ref{S_H}), to quadratic order in $k$, reads
\begin{align}
H\approx-\frac{\hbar^2(v^2+v_4^2)}{\gamma_1}\left(\begin{array}{cc}
0&(\xi k_x-ik_y)^2\\(\xi k_x+ik_y)^2&0\end{array}\right)\nonumber\\
+\frac{U}2\left(\begin{array}{cc}
-1&0\\0&1\end{array}\right)+\frac{2vv_4\hbar^2k^2}{\gamma_1}\left(\begin{array}{cc}
1&0\\0&1\end{array}\right),
\end{align}
which, apart from the small correction $v_4^2$, coincides with the effective low-energy Hamiltonian \cite{McCann2013} taking into account electron chirality, gap opening, and the electron-hole asymmetry correction $\propto v_4$. On the other hand, the eigenvalues of (\ref{S_H})
\begin{equation}
E_\pm=\frac{E_+^0+E_-^0}2\pm\sqrt{\left(\frac{E_+^0-E_-^0}2
\right)^2+\left(\frac{U}2\right)^2}\label{S_En}
\end{equation}
at high momenta $k$ tend to correct linear-dispersing dependencies $E_\pm\approx(\pm v+v_4)\hbar k$ with the electron-hole asymmetry. Numerical checks justify that the approximate Hamiltonian (\ref{S_H}) provides electron dispersions and wave functions which are close to those obtained from the full $(4\times4)$ Hamiltonian \cite{McCann2013} within 10\% at carrier densities $|n|\leq10^{13}\,\mbox{cm}^{-2}$ and interlayer potential differences $|U|\leq0.15\,\mbox{eV}$.

For approximation of noninteracting electrons, the chemical potentials $\mu_i$ ($i=1,2$) entering the electrostatic equations (\ref{el16})-(\ref{el19}) are electron energies (\ref{S_En}) at the Fermi momenta $k_{\mathrm{F}i}$:
\begin{equation}
\mu_i=\left.\frac{E_+^0+E_-^0}2+\mathrm{sgn}(n_i)\sqrt{\left(\frac{E_+^0-E_-^0}2
\right)^2+\left(\frac{U_i}2\right)^2}\right|_{k=k_{\mathrm{F}i}}.\label{S_mu_i}
\end{equation}
The sign of charge carrier density $n_i$ determines whether the Fermi level is located in the valence band ($n_i<0$) or the conduction band ($n_i>0$). The carrier density itself is related to the Fermi momentum as
\begin{equation}
k_{\mathrm{F}i}=\sqrt{\pi|n_i|}.\label{S_k_F}
\end{equation}
The sublayer density imbalance $\Delta n_i$ arises at $U_i\neq0$ due to unequal population of different sublattices. In the simplest model of BLG with quadratic electron-hole symmetric dispersion, the imbalance is
\begin{equation}
\Delta n_i=\frac{U_in_\perp}{2\gamma_1}\log\left(\frac{|n_i|}{2n_\perp}+\frac12\sqrt{\left(\frac{n_i}{n_\perp}\right)^2+\left(\frac{U_i}{2\gamma_1}\right)^2}\right),\label{S_Delta_n_i}
\end{equation}
where $n_\perp=\gamma_1^2/\pi\hbar^2v^2$. Taking into account more complicated dispersions (\ref{S_En}) would affect only the expression under the logarithm providing a small correction. Thus we use the simple equation Eq.~(\ref{S_Delta_n_i}) in order to relate the density imbalances $\Delta n_i$ to $n_i$ and $U_i$.

\subsection{Tunneling current}

The tunneling current flowing from top to bottom BLG can be calculated using the Fermi's golden rule as
\begin{align}
I\propto-\sum_{\mathbf{k}\gamma_1\gamma_2}M_{\mathbf{k}\gamma_1\gamma_2}\int dE&\left[n_\mathrm{F}(E+eV_\mathrm{b})-n_\mathrm{F}(E)\right]\nonumber\\
&\times\rho(E+\mu_1+eV_\mathrm{b}-E_{\mathbf{k}+\Delta\mathbf{K},\gamma_1})\rho(E+\mu_2-E_{\mathbf{k}\gamma_2}).\label{S_I}
\end{align}
Here $n_\mathrm{F}(E)=[e^{E/T}+1]^{-1}$ is the Fermi-Dirac distribution, $\rho(E)=(\sqrt{2\pi}\Gamma)^{-1}\exp(-E^2/2\Gamma^2)$ is the Gaussian spectral function of broadened electron states with the energy width $\Gamma$, $M_{\mathbf{k}\gamma_1\gamma_2}$ is the square modulus of tunneling matrix element connecting electron state with momentum $\mathbf{k}+\Delta\mathbf{K}$ and band $\gamma_1=\pm1$ in the top BLG with electron state with momentum $\mathbf{k}$ and band $\gamma_2=\pm1$ in the bottom BLG. The energies of these states $E_{\mathbf{k}+\Delta\mathbf{K},\gamma_1}$ and $E_{\mathbf{k}\gamma_2}$ measured from the Dirac points of the corresponding BLGs are given by Eq.~(\ref{S_En}). The relative momentum shift of electron states is connected to the twist angle $\theta$ as
\begin{equation}
\Delta K\approx\frac{4\pi\theta}{3a},
\end{equation}
where $a=2.46\,\mbox{\AA}$ is the graphene lattice constant.

The energy $E$ in Eq.~(\ref{S_I}) is defined in such a way that the Fermi level of the bottom BLG is $E=0$, then the Fermi level of the top BLG is located at $E=-eV_\mathrm{b}$. The difference $n_\mathrm{F}(E+eV_\mathrm{b})-n_\mathrm{F}(E)$ of Fermi-Dirac distributions means that the tunneling current is given by pairs of electron states which are occupied in the top BLG and unoccupied in the bottom BLG (or vice versa, in this case the current sign is reversed). This difference is multiplied by the square modulus of tunneling matrix element $M_{\mathbf{k}\gamma_1\gamma_2}$ and by the product of densities of states in top and bottom BLGs, where electron state energies, counted from the bottom BLG Fermi level, would be $E_{\mathbf{k}+\Delta\mathbf{K},\gamma_1}-\mu_1-eV_\mathrm{b}$ and $E_{\mathbf{k}\gamma_2}-\mu_2$ in the absence of broadening. In the presence of broadening, we obtain the Gaussians centered around these energies.

In our model, we assume that the tunneling occurs only between the closest sublattices, i.e. between sublattice A1 of top BLG and sublattice B2 of bottom BLG. By this reason, the tunneling matrix element should take into account overlap of electron wave functions at these sublattices:
\begin{equation}
M_{\mathbf{k}\gamma_1\gamma_2}=\left|(\Psi_{\mathbf{k}\gamma_2})^*_\mathrm{B2}(\Psi_{\mathbf{k}+\Delta\mathbf{K},\gamma_1})_\mathrm{A1}\right|^2,
\end{equation}
where $\Psi_{\mathbf{k}\gamma}$ is the eigenvector (two-component column) of the Hamiltonian (\ref{S_H}) corresponding to the eigenvalue $E_\gamma$ (\ref{S_En}) with the momentum $k$.

Thus, numerical solution of the electrostatic equations (\ref{el16})-(\ref{el19}) together with Eqs.~(\ref{S_mu_i})-(\ref{S_Delta_n_i}) allows to find the tunneling current (\ref{S_I}) at given voltages $V_\mathrm{tg}$, $V_\mathrm{bg}$, and $V_\mathrm{b}$.

\subsection{Calculation parameters}

Similarly to our previous paper \cite{Sokolik2025}, we use the following generally accepted \cite{McCann2013} parameters of electron dispersion in BLG:
\begin{equation}
\gamma_1=0.381\,\mbox{eV},\qquad v=10^6\,\mbox{m/s},\qquad v_4=0.05v.
\end{equation}
Our calculation of the tunneling current using Eq.~(\ref{S_I}) are carried out at temperature $T=4.5\,\mbox{K}$, and energy width $\Gamma=30\,\mbox{K}$ for Device 1 and $\Gamma=10\,\mbox{K}$ for Device 2. The parameters of electrostatic model are fitted to reproduce the experimental data as close as possible:
\begin{align}
&\mbox{Device 1:}\qquad d_\mathrm{t}=37.75\,\mbox{nm},\qquad
d_\mathrm{m}=2\,\mbox{nm},\qquad d_\mathrm{b}=234.8\,\mbox{nm};\\
&\mbox{Device 2:}\qquad d_\mathrm{t}=40.4\,\mbox{nm},\qquad
d_\mathrm{m}=1.5\,\mbox{nm},\qquad d_\mathrm{b}=47.8\,\mbox{nm}.
\end{align}
The distance between graphene sublayers of BLG is $d_0=3.35\,\mbox{\AA}$, and all dielectric constants are assumed to be equal to that for bulk hBN: $\varepsilon_\mathrm{t}=\varepsilon_\mathrm{m}=\varepsilon_\mathrm{b}=3.2$. The increased value of effective back gate dielectric spacer thickness $d_\mathrm{b}$ for Device 1 is caused by the presence of additional 290-nm thick $\mathrm{SiO}_2$ layer above the bottom gate.

\section{Additional maps of tunneling current}

\begin{figure}[!p]         
\centering
\includegraphics[width=1\textwidth]{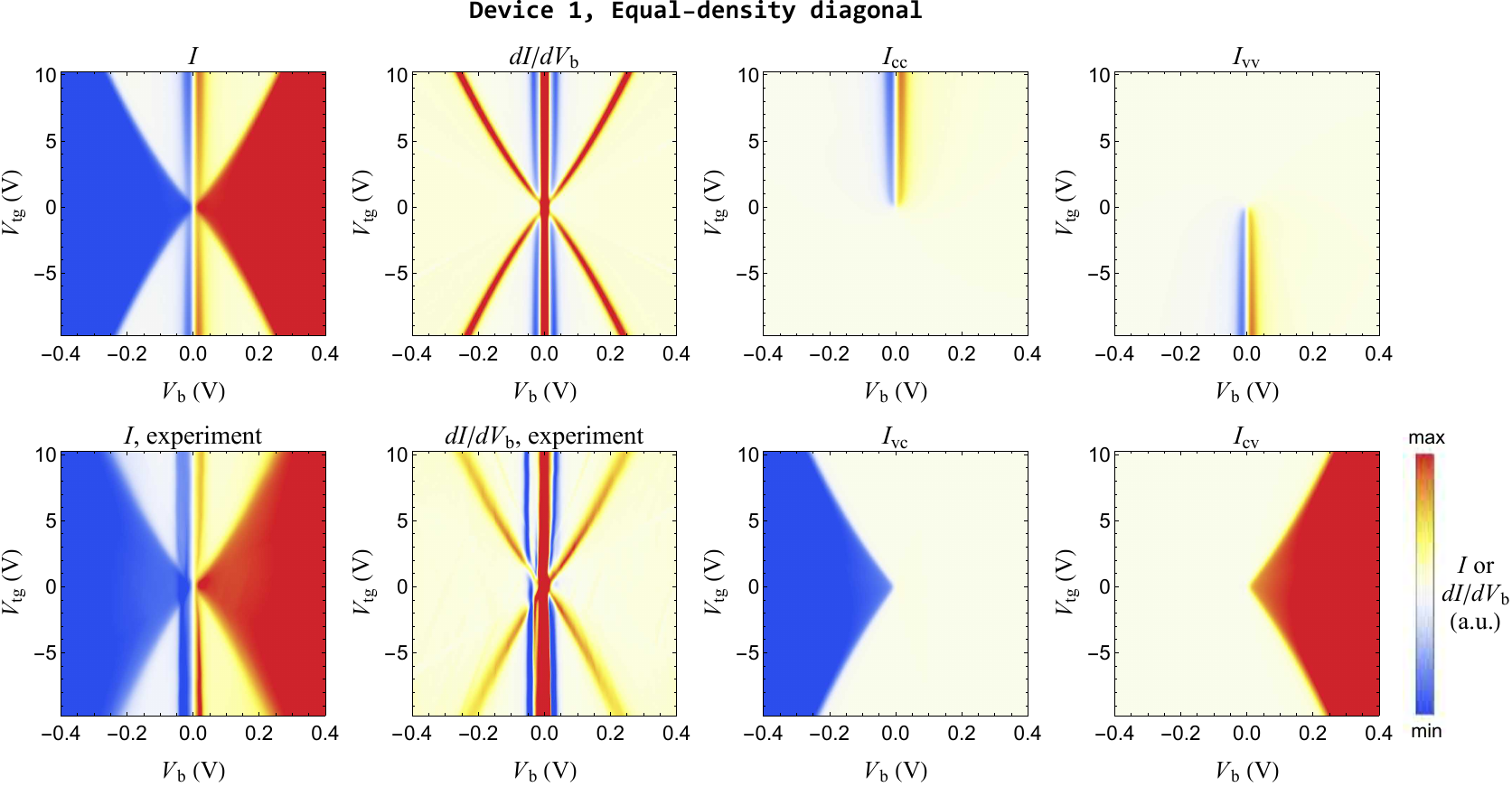}
\caption{2D maps for tunneling current in Device 1 and its derivative versus bias voltage $V_\mathrm{b}$ and $V_\mathrm{tg}$, when the latter changes simultaneously with $V_\mathrm{bg}$ to satisfy the condition of equal carrier densities in both BLGs at $V_\mathrm{b}=0$. The calculated current is resolved over combinations of initial and final bands of top and bottom BLGs (right panels).}
\label{FigST4f2}
\end{figure}

\begin{figure}[!p]         
\centering
\includegraphics[width=1\textwidth]{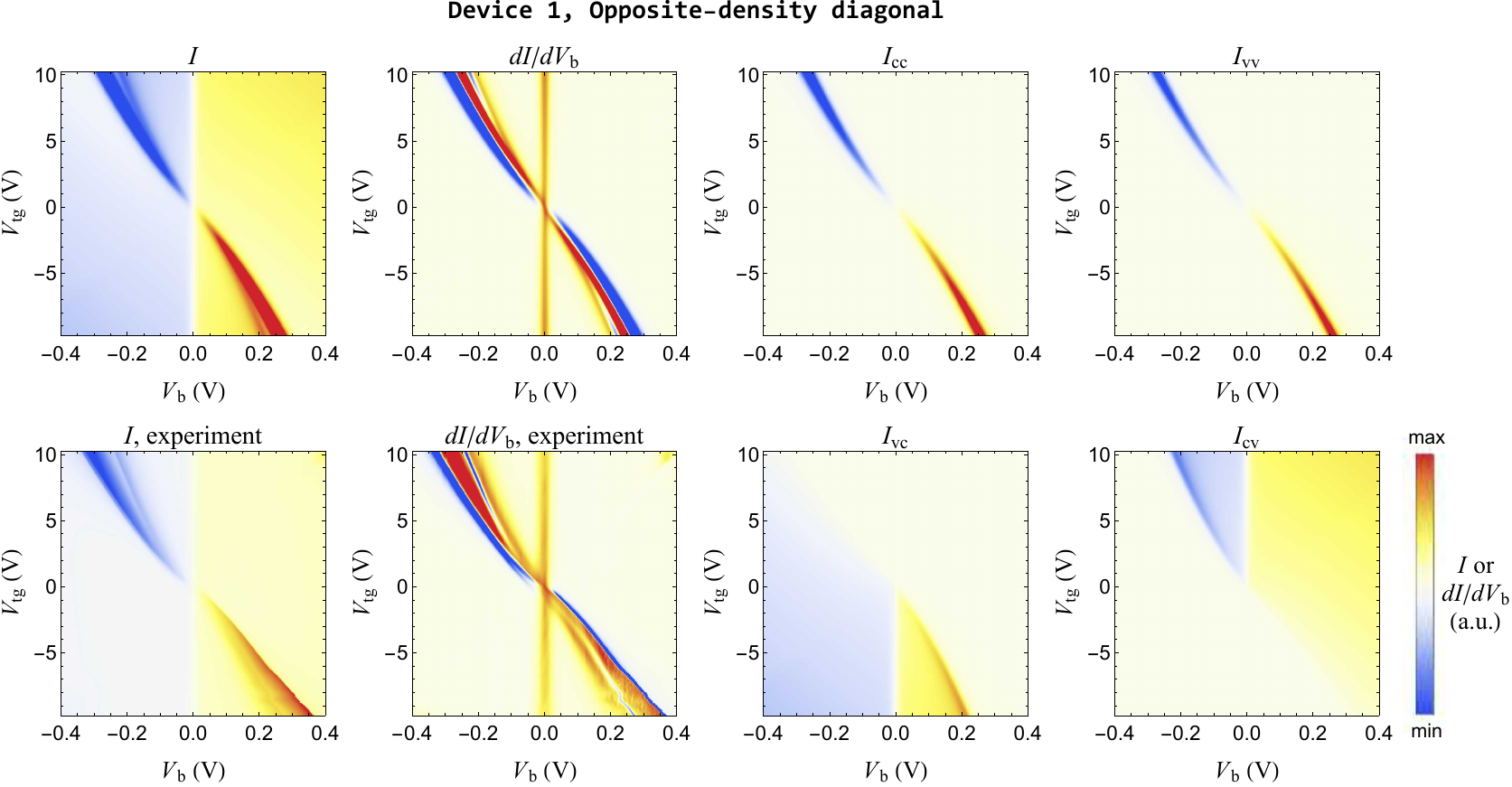}
\caption{The same as Fig.~\ref{FigST4f2} but for $V_\mathrm{tg}$ changing simultaneously with $V_\mathrm{bg}$ to satisfy the condition of opposite carrier densities in two BLGs.}
\label{FigST4f1}
\end{figure}

\begin{figure}[!p]         
\centering
\includegraphics[width=1\textwidth]{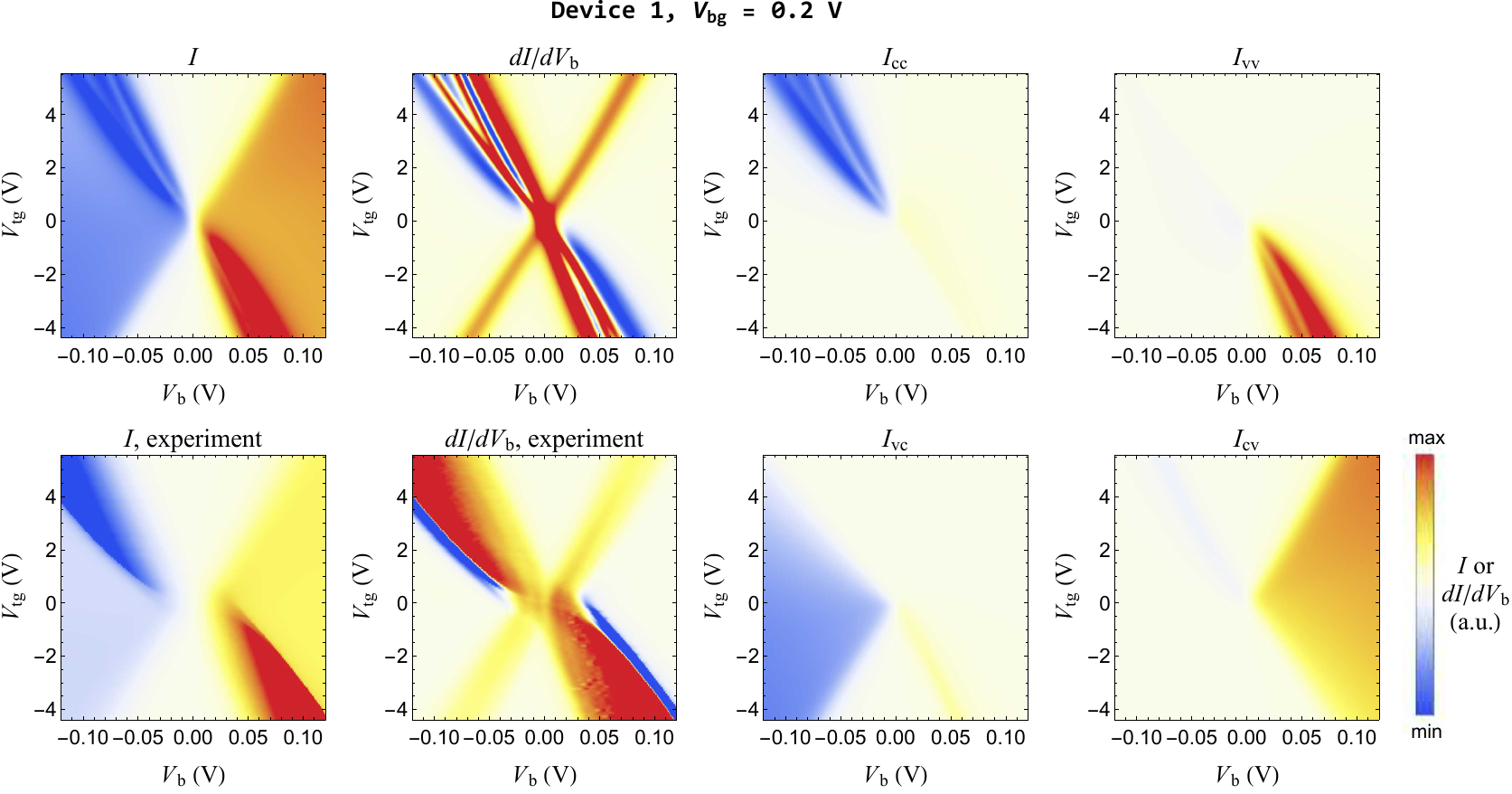}
\caption{2D maps for tunneling current and its derivative in Device 1 versus bias voltage $V_\mathrm{b}$ and $V_\mathrm{tg}$ at fixed $V_\mathrm{bg}=0.2\,\mbox{V}$.}
\label{FigST4f3}
\end{figure}

\begin{figure}[!p]         
\centering
\includegraphics[width=1\textwidth]{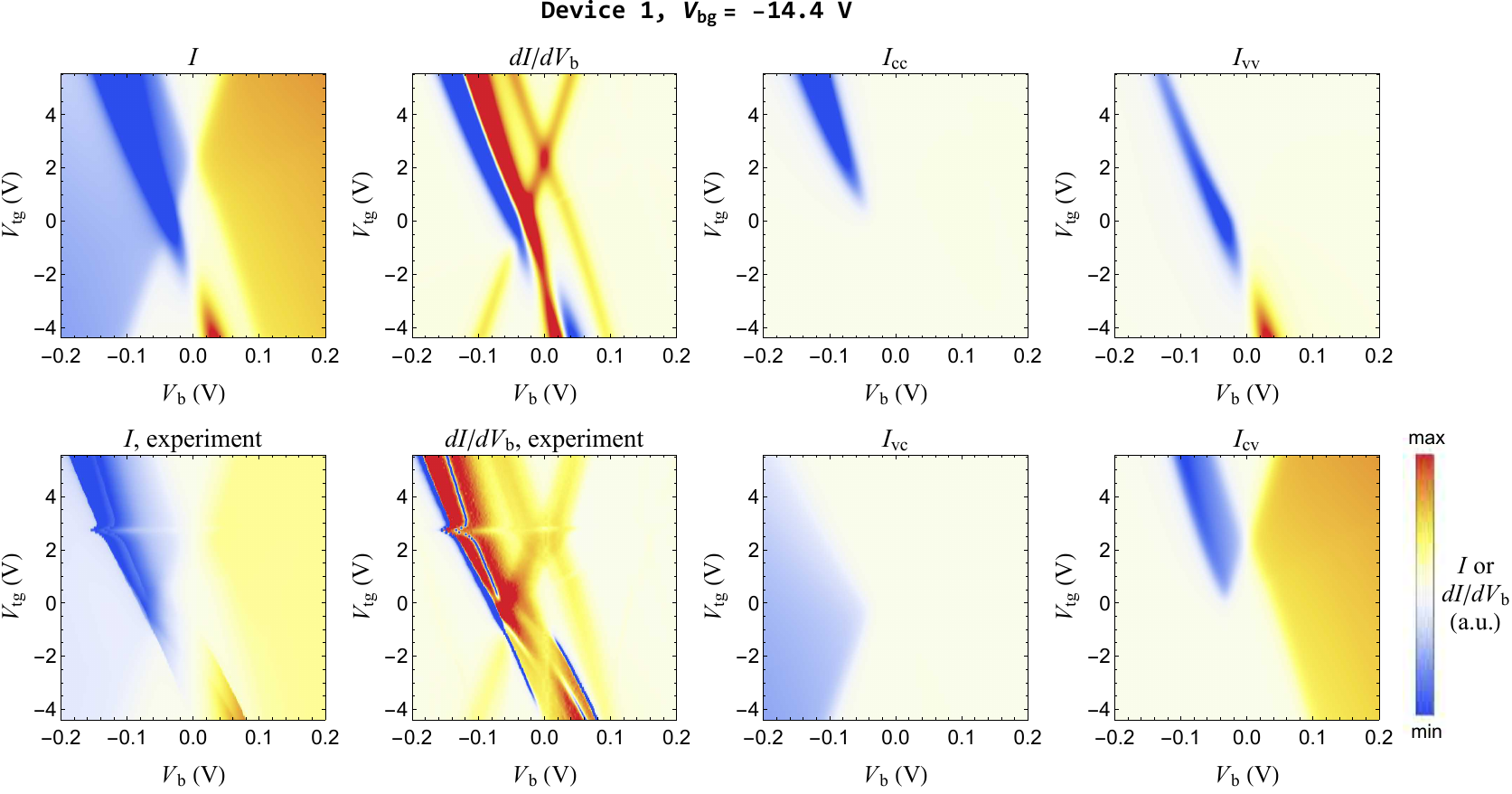}
\caption{2D maps for tunneling current and its derivative in Device 1 versus bias voltage $V_\mathrm{b}$ and $V_\mathrm{tg}$ at fixed $V_\mathrm{bg}=-14.4\,\mbox{V}$.}
\label{FigST4f4}
\end{figure}

\begin{figure}[!p]         
\centering
\includegraphics[width=1\textwidth]{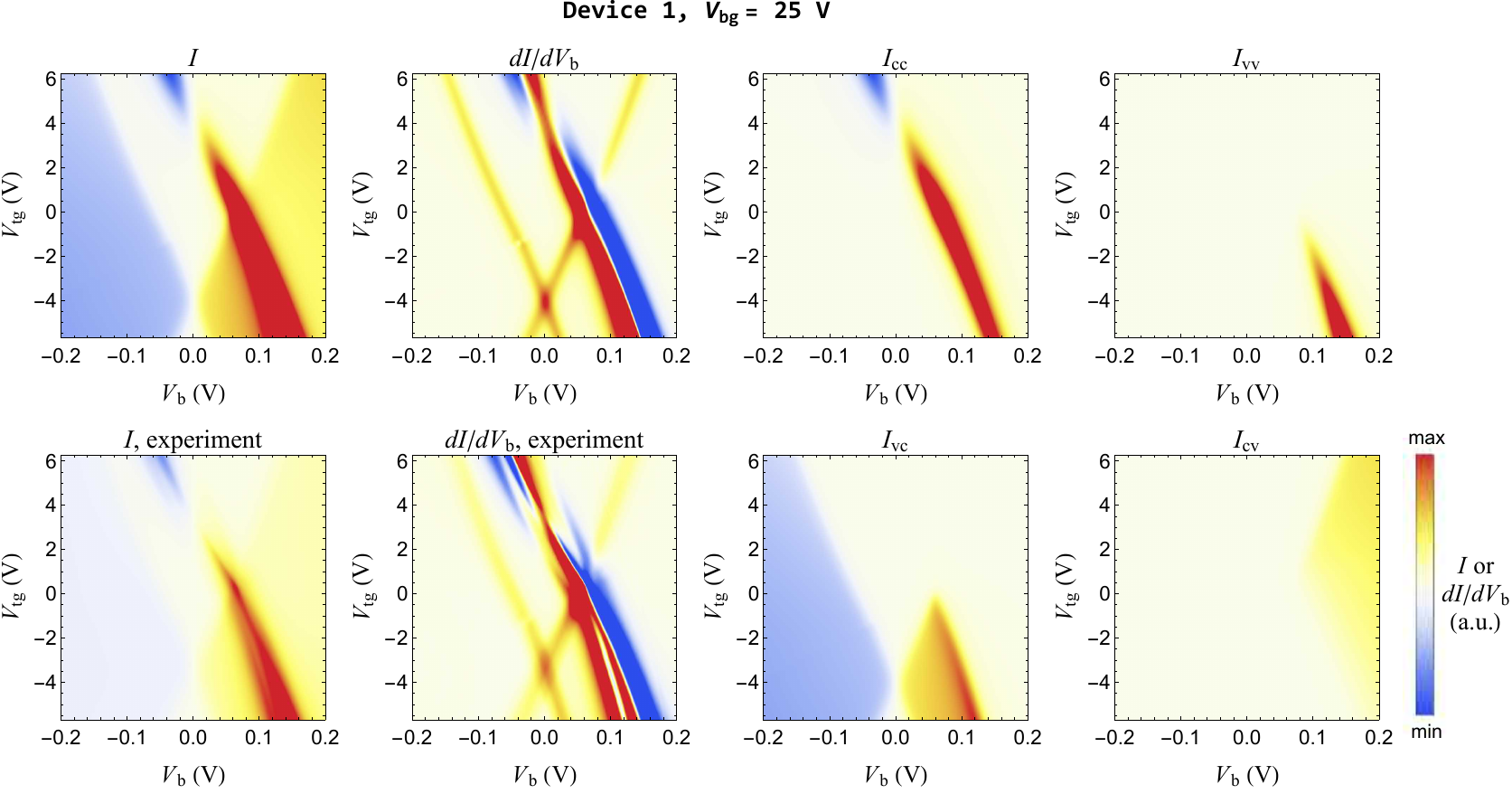}
\caption{2D maps for tunneling current and its derivative in Device 1 versus bias voltage $V_\mathrm{b}$ and $V_\mathrm{tg}$ at fixed $V_\mathrm{bg}=25\,\mbox{V}$.}
\label{FigST4f5}
\end{figure}

\begin{figure}[!p]         
\centering
\includegraphics[width=1\textwidth]{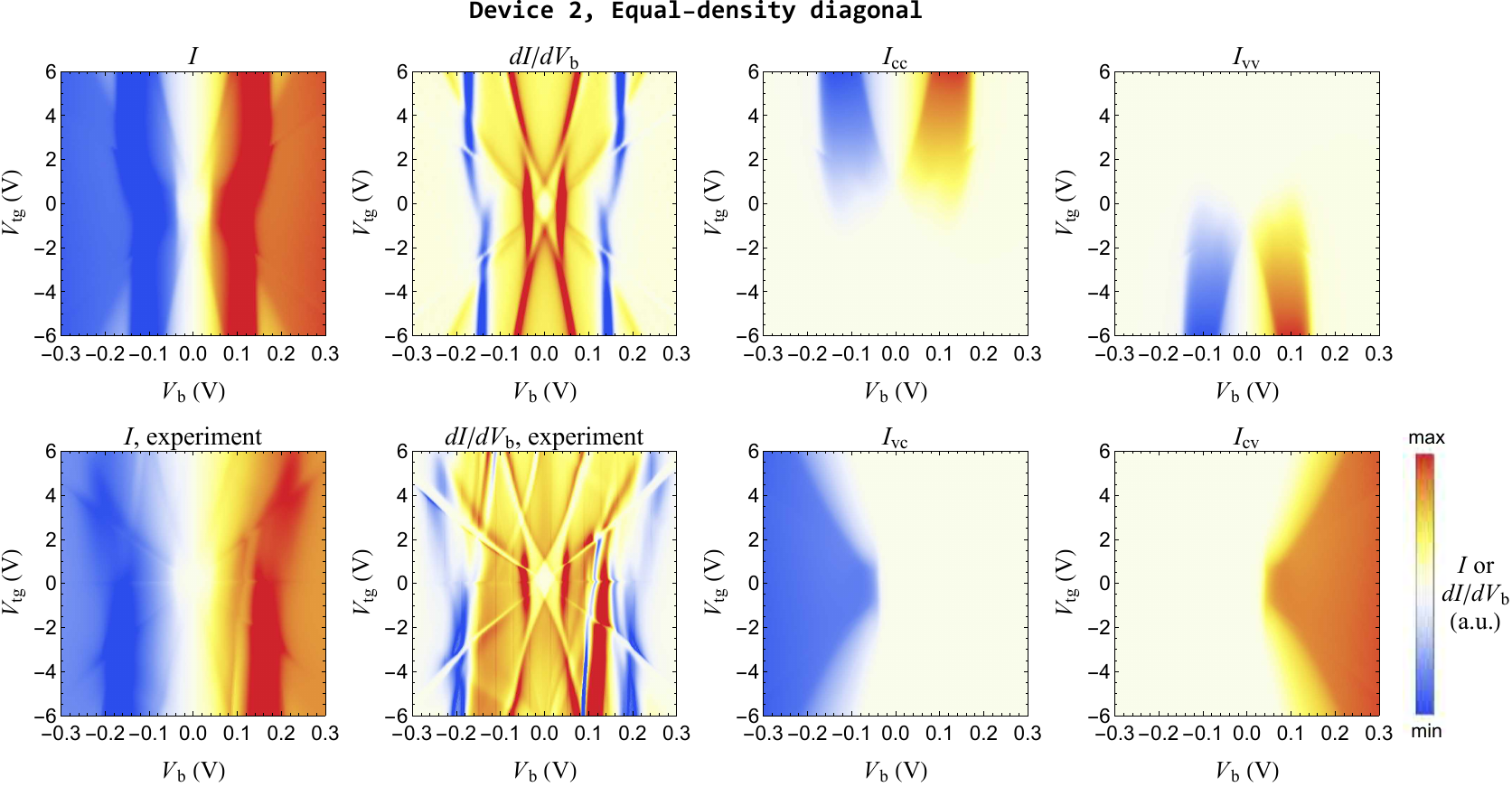}
\caption{2D maps for tunneling current and its derivative in Device 2 versus bias voltage $V_\mathrm{b}$ and $V_\mathrm{tg}$, when the latter changes simultaneously with $V_\mathrm{bg}$ to satisfy the condition of equal carrier densities in both BLGs at $V_\mathrm{b}=0$.}
\label{FigST5f2}
\end{figure}

\begin{figure}[!p]         
\centering
\includegraphics[width=1\textwidth]{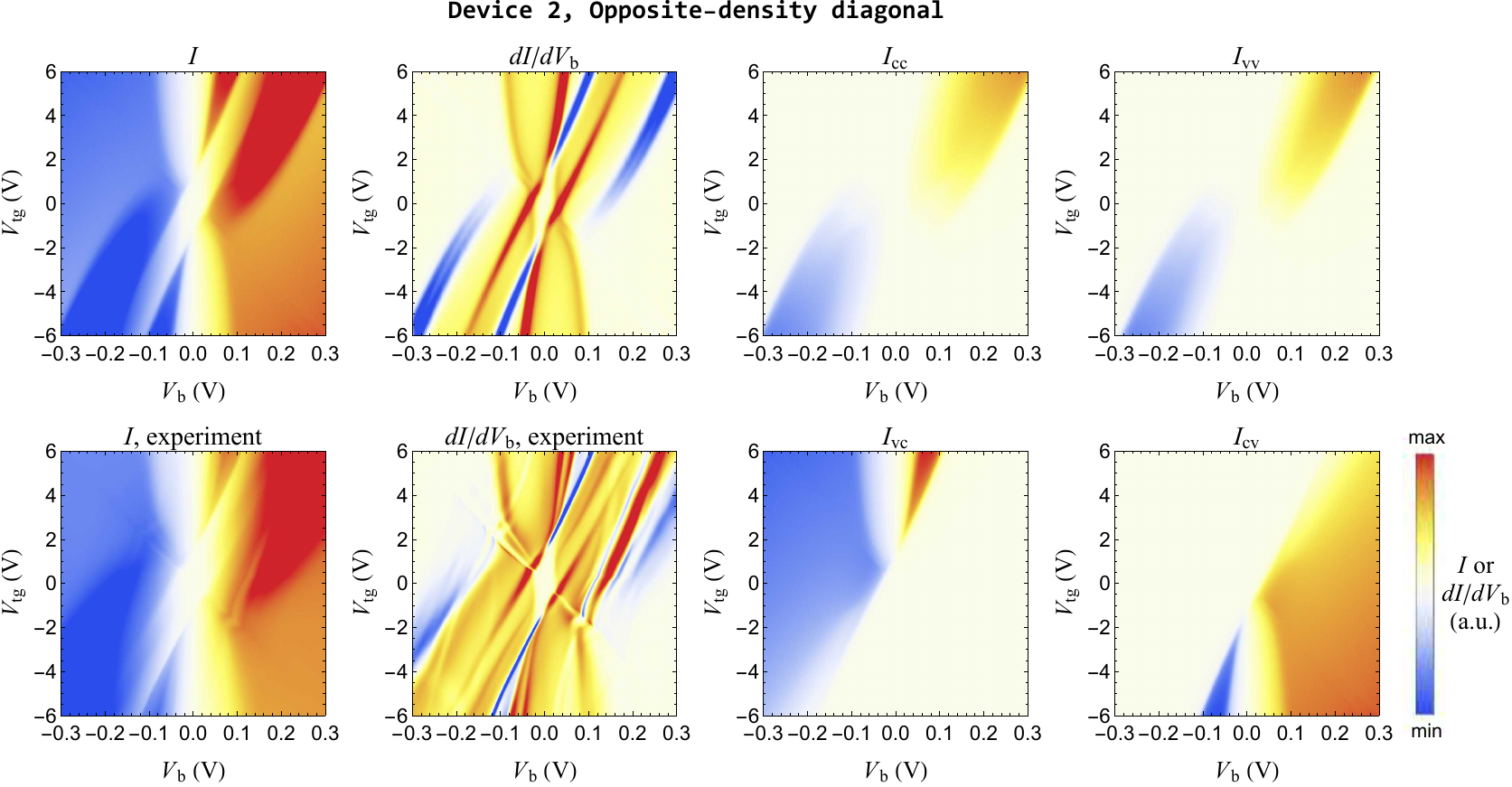}
\caption{The same as Fig.~\ref{FigST5f1} but for $V_\mathrm{tg}$ changing simultaneously with $V_\mathrm{bg}$ to satisfy the condition of opposite carrier densities in two BLGs.}
\label{FigST5f1}
\end{figure}

\begin{figure}[!p]         
\centering
\includegraphics[width=1\textwidth]{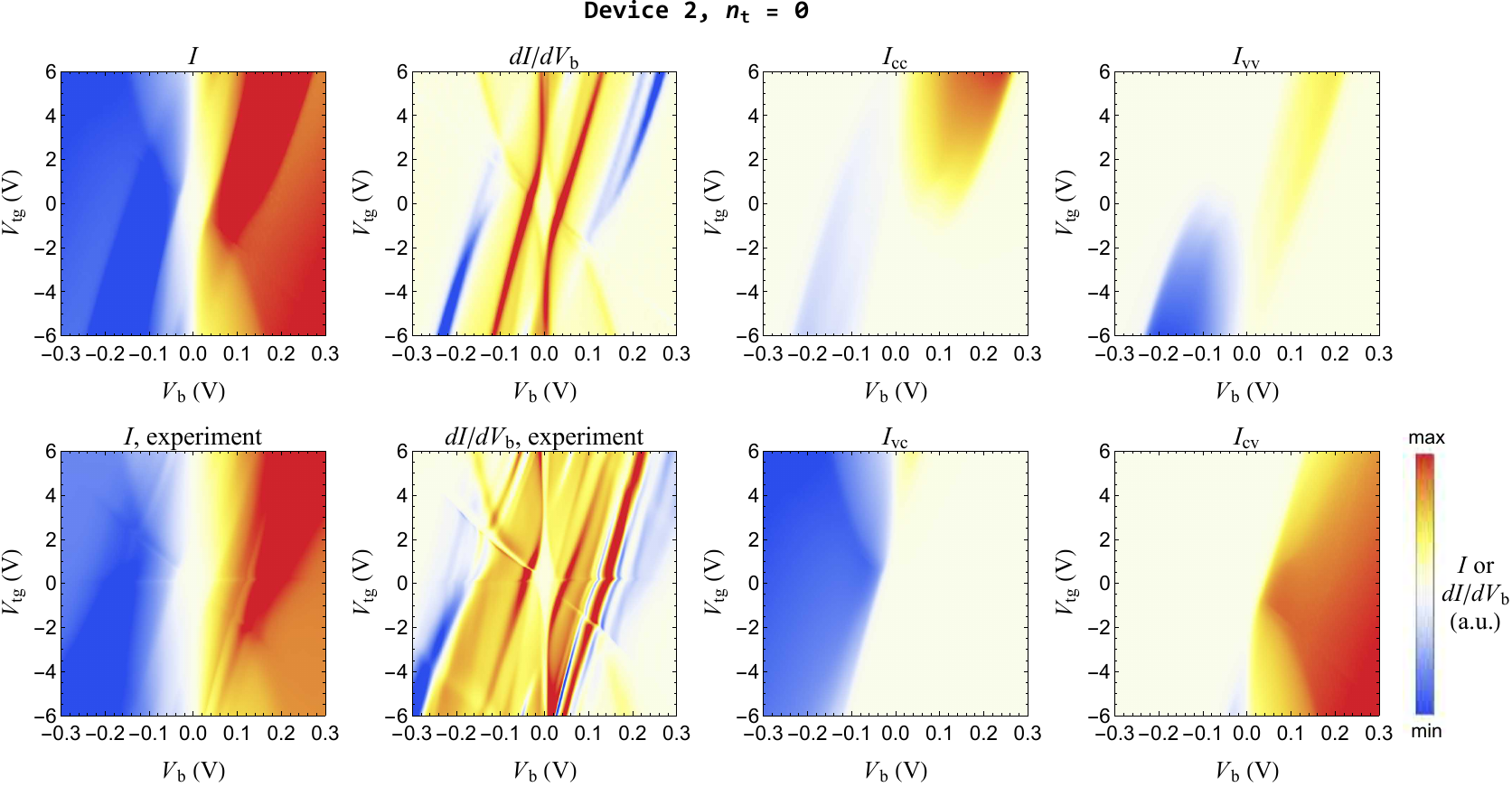}
\caption{2D maps for tunneling current and its derivative in Device 2 versus bias voltage $V_\mathrm{b}$ and $V_\mathrm{tg}$, when the latter changes simultaneously with $V_\mathrm{bg}$ to satisfy the condition of zero carrier density (charge neutrality) $n_\mathrm{t}=0$ of top BLG.}
\label{FigST5f3}
\end{figure}

In this section we present additional, more detailed 2D maps of the tunneling current $I$ and its derivative $dI/dV_\mathrm{b}$, both experimentally measured and calculated theoretically, see Figures~\ref{FigST4f2}-\ref{FigST4f5} for Device 1 and Figures~\ref{FigST5f2}-\ref{FigST5f3} for Device 2. The calculated current is resolved by initial (in top BLG) and final (in bottom BLG) electron band: for example, the partial current $I_\mathrm{cv}$ corresponds to the tunneling from conduction band of top BLG to valence band of bottom BLG.

Figures~\ref{FigST4f2} and \ref{FigST5f2} demonstrate how the tunneling characteristics of Device 1 and Device 2 respectively change along the equal-density diagonal, i.e. when the doping levels of both BLGs are equal at $V_\mathrm{b}=0$. In addition to the main text, here we show the calculated and measured differential conductance $dI/dV_\mathrm{b}$, which highlights sharp features of tunneling characteristics, such as peaks, their splittings, and steps. Resolution of the calculated current by both initial and final bands demonstrates how their contributions depend on polarity of $V_\mathrm{b}$ and on the signs of carrier densities. Figures~\ref{FigST4f1} and \ref{FigST5f1} present similar picture, but for opposite-density diagonal, when two BLGs are doped with opposite carrier densities.

Figures~\ref{FigST4f3}-\ref{FigST4f5} show tunneling maps for Device 1 at fixed values of $V_\mathrm{bg}$, and Figure~\ref{FigST5f3} show the tunneling map for Device 2, where $V_\mathrm{tg}$ and $V_\mathrm{bg}$ are changed simultaneously to provide charge neutrality of top BLG at $V_\mathrm{b}=0$. These maps demonstrate the same features as the diagonal maps discussed in the main text: steps of the unlike-band currents ($I_\mathrm{cv}$ and $I_\mathrm{vc}$) and resonant peaks of the like-band currents ($I_\mathrm{cc}$ and $I_\mathrm{vv}$), which become more diffuse in Device 2 due to momentum mismatch of electron dispersions.

\section{Additional I-V characteristics}

\begin{figure}[!p]         
\centering
\includegraphics[width=1\textwidth]{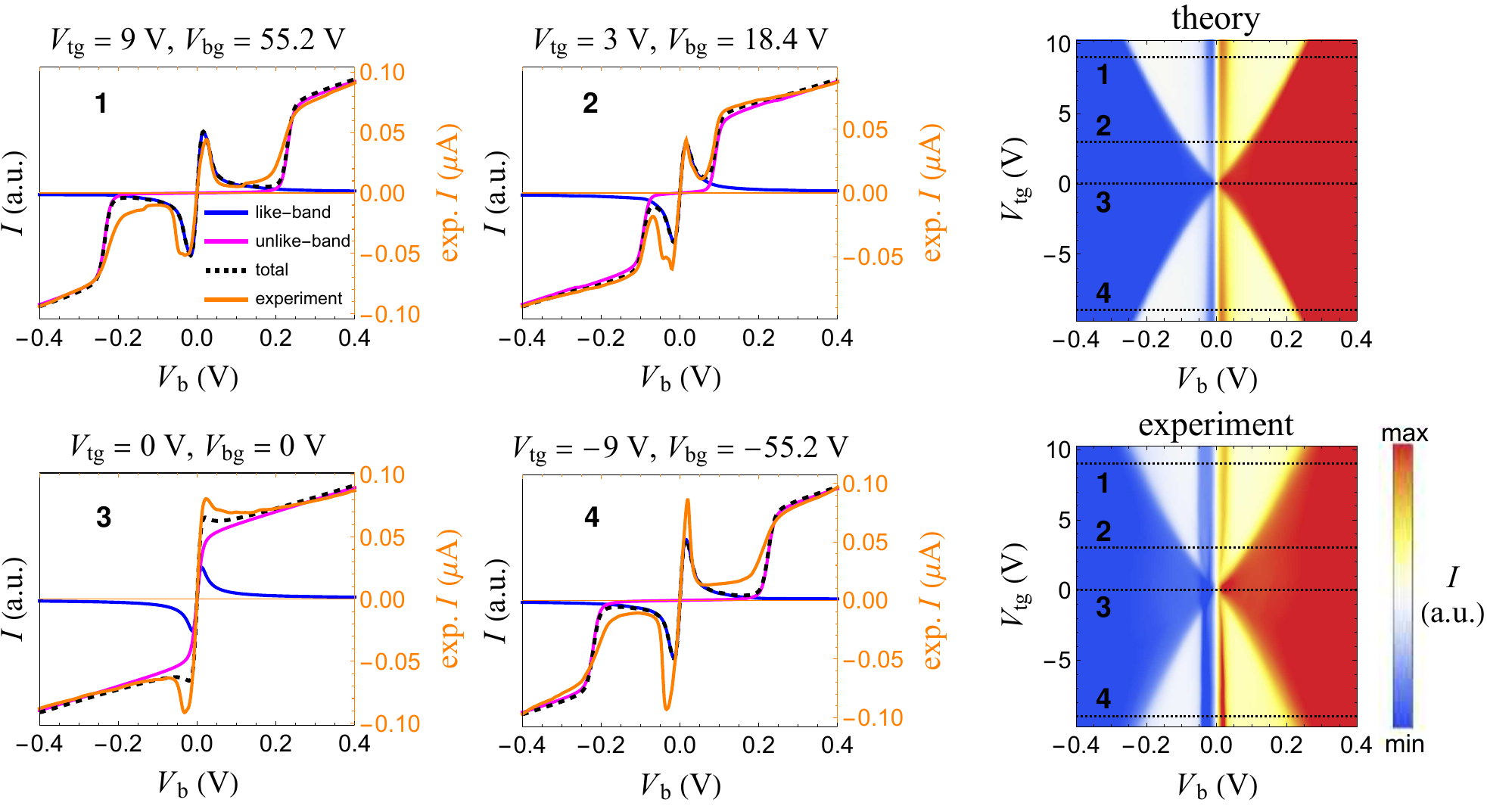}
\caption{Tunneling characteristics for Device 1 at different points $(V_\mathrm{tg},V_\mathrm{bg})$ of the equal-density diagonal. These characteristics are cross-sections (numbered by 1-4) of 2D tunneling current maps shown in right panels, which are the same as in Figure~\ref{FigST4f2}.}
\label{FigST4cf2}
\end{figure}

\begin{figure}[!p]         
\centering
\includegraphics[width=1\textwidth]{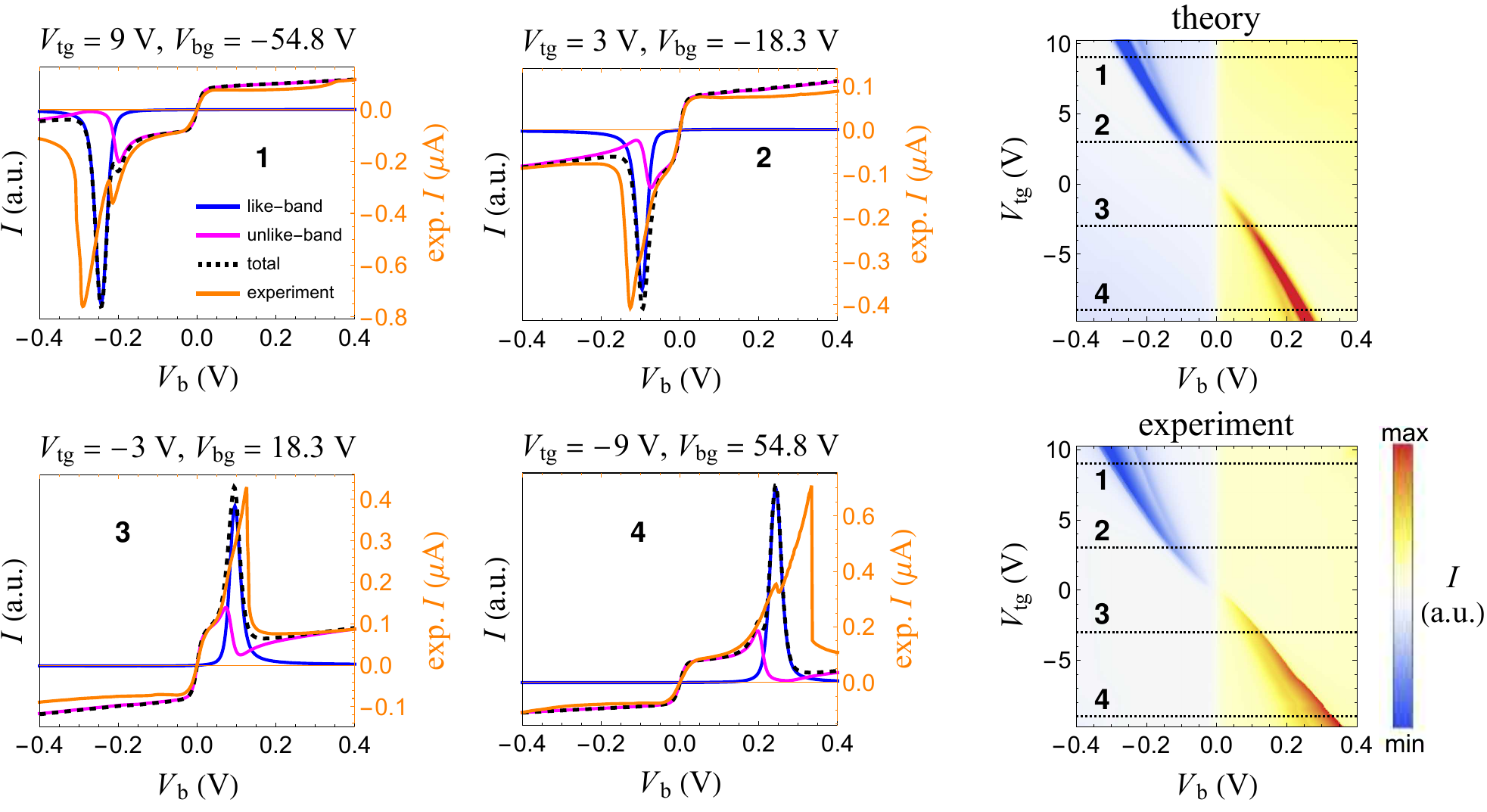}
\caption{The same as Figure~\ref{FigST4cf2}, but the points $(V_\mathrm{tg},V_\mathrm{bg})$ are taken at the opposite-density diagonal. 2D tunneling current maps in right panels are the same as in Figure~\ref{FigST4f1}.}
\label{FigST4cf1}
\end{figure}

\begin{figure}[!p]         
\centering
\includegraphics[width=1\textwidth]{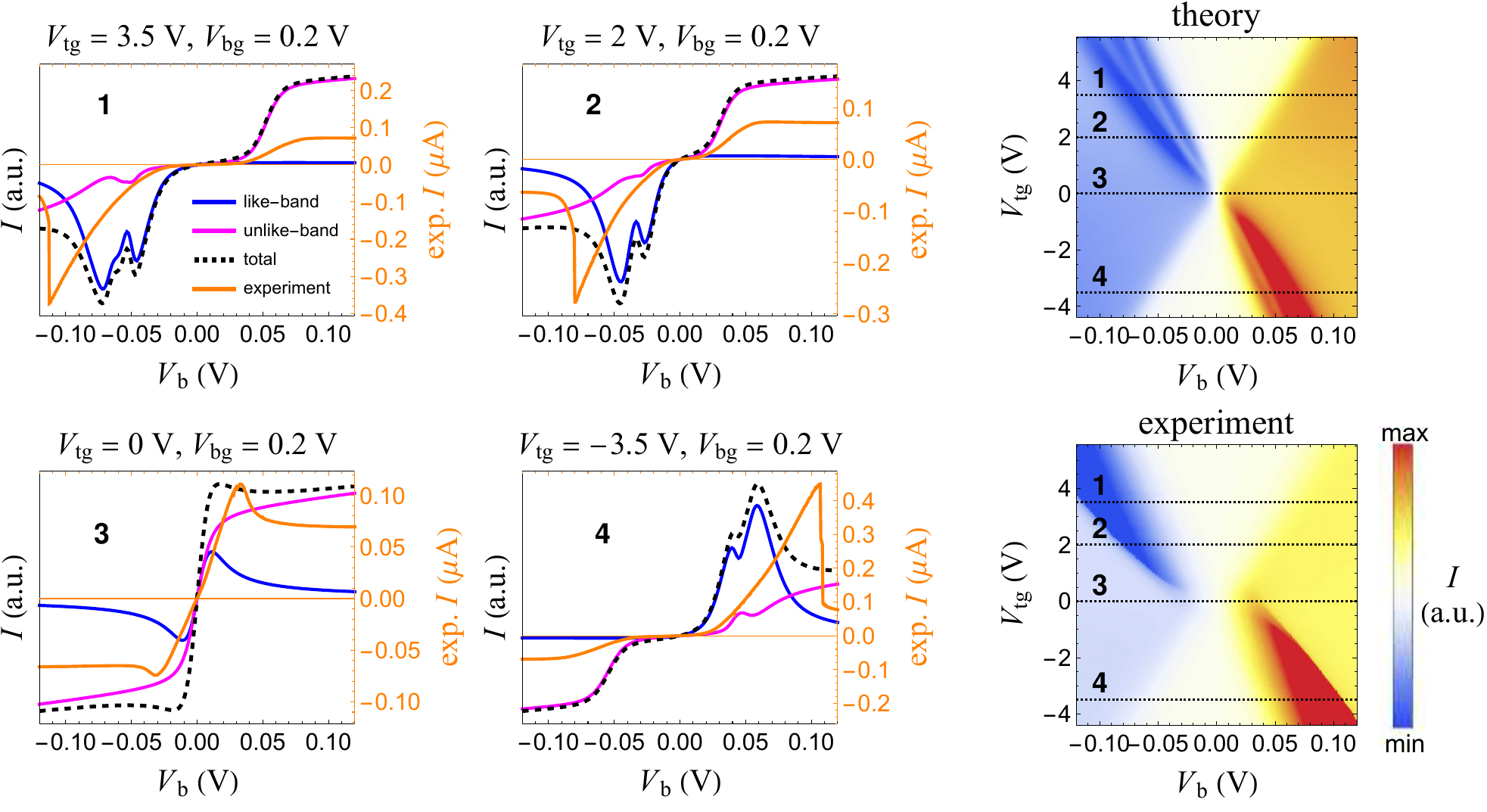}
\caption{Tunneling characteristics for Device 1 at different points values of $V_\mathrm{tg}$ at fixed $V_\mathrm{bg}=0.2\,\mbox{V}$. These characteristics are cross-sections (numbered by 1-4) of 2D tunneling current maps shown in right panels, which are the same as in Figure~\ref{FigST4f3}.}
\label{FigST4cf3}
\end{figure}

\begin{figure}[!p]         
\centering
\includegraphics[width=1\textwidth]{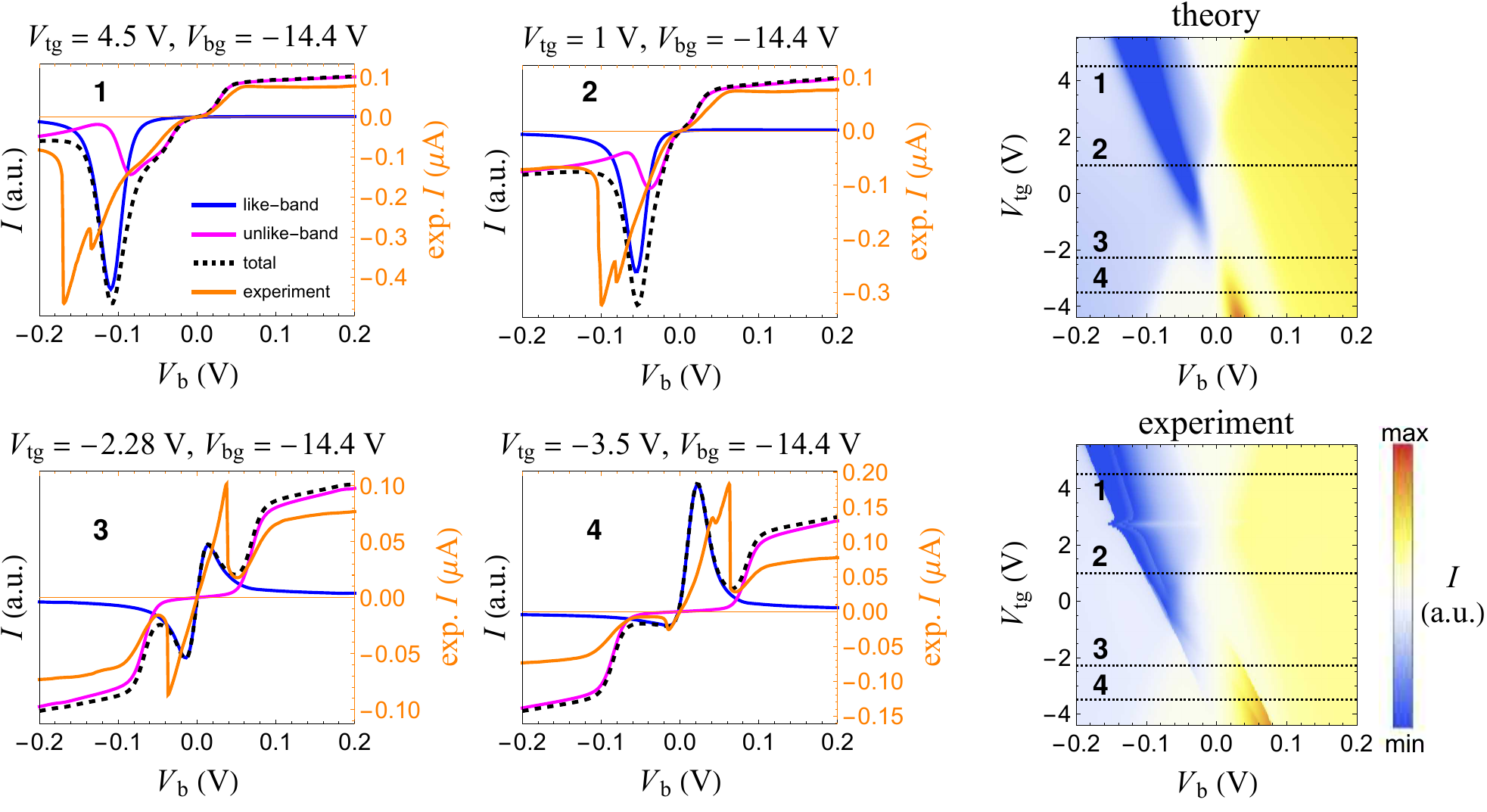}
\caption{The same as Figure~\ref{FigST4cf3}, but for $V_\mathrm{bg}=-14.4\,\mbox{V}$. 2D tunneling current maps in right panels are the same as in Figure~\ref{FigST4f4}.}
\label{FigST4cf4}
\end{figure}

\begin{figure}[!p]         
\centering
\includegraphics[width=1\textwidth]{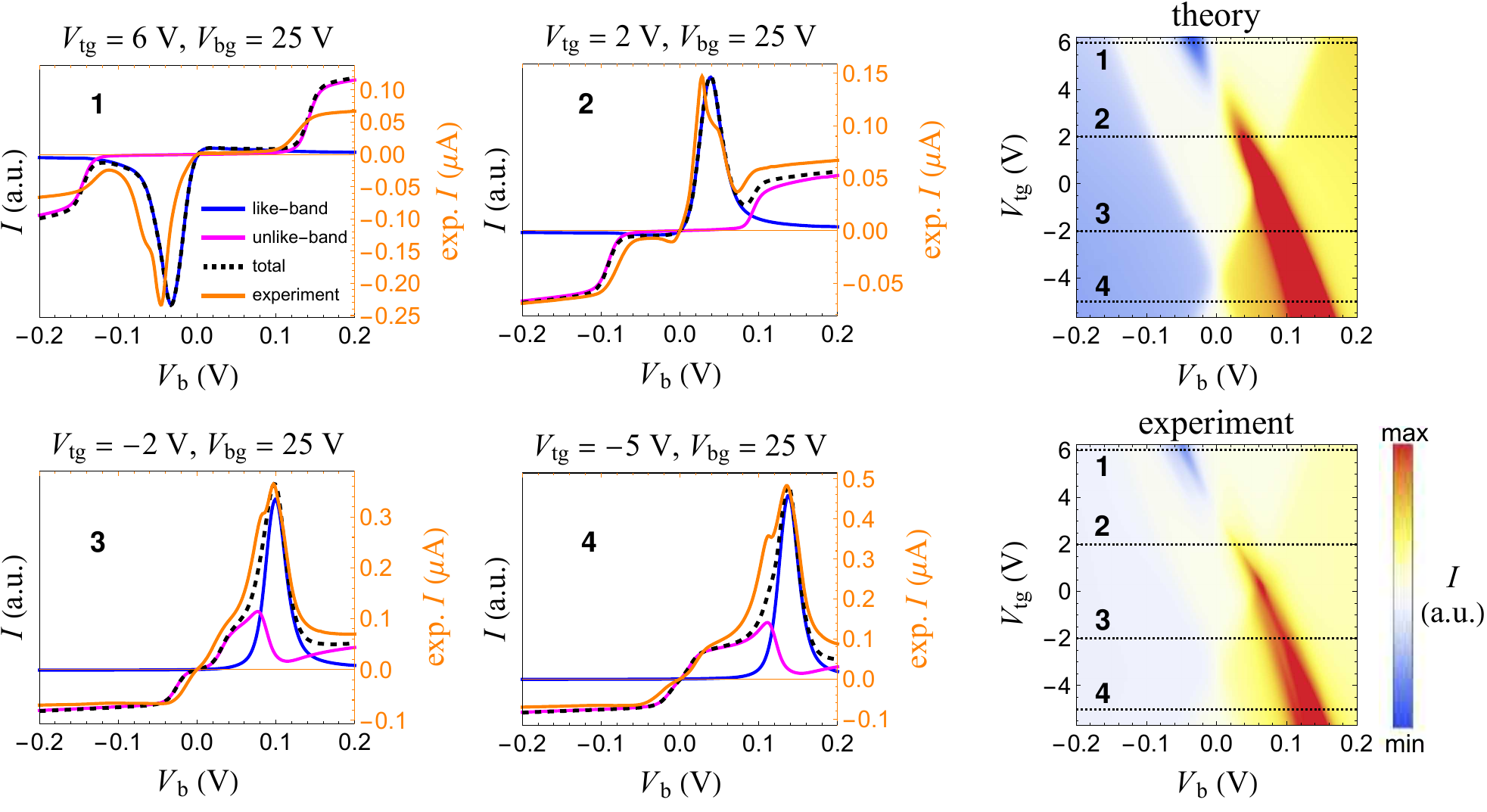}
\caption{The same as Figure~\ref{FigST4cf3}, but for $V_\mathrm{bg}=25\,\mbox{V}$. 2D tunneling current maps in right panels are the same as in Figure~\ref{FigST4f5}.}
\label{FigST4cf5}
\end{figure}

\begin{figure}[!p]         
\centering
\includegraphics[width=1\textwidth]{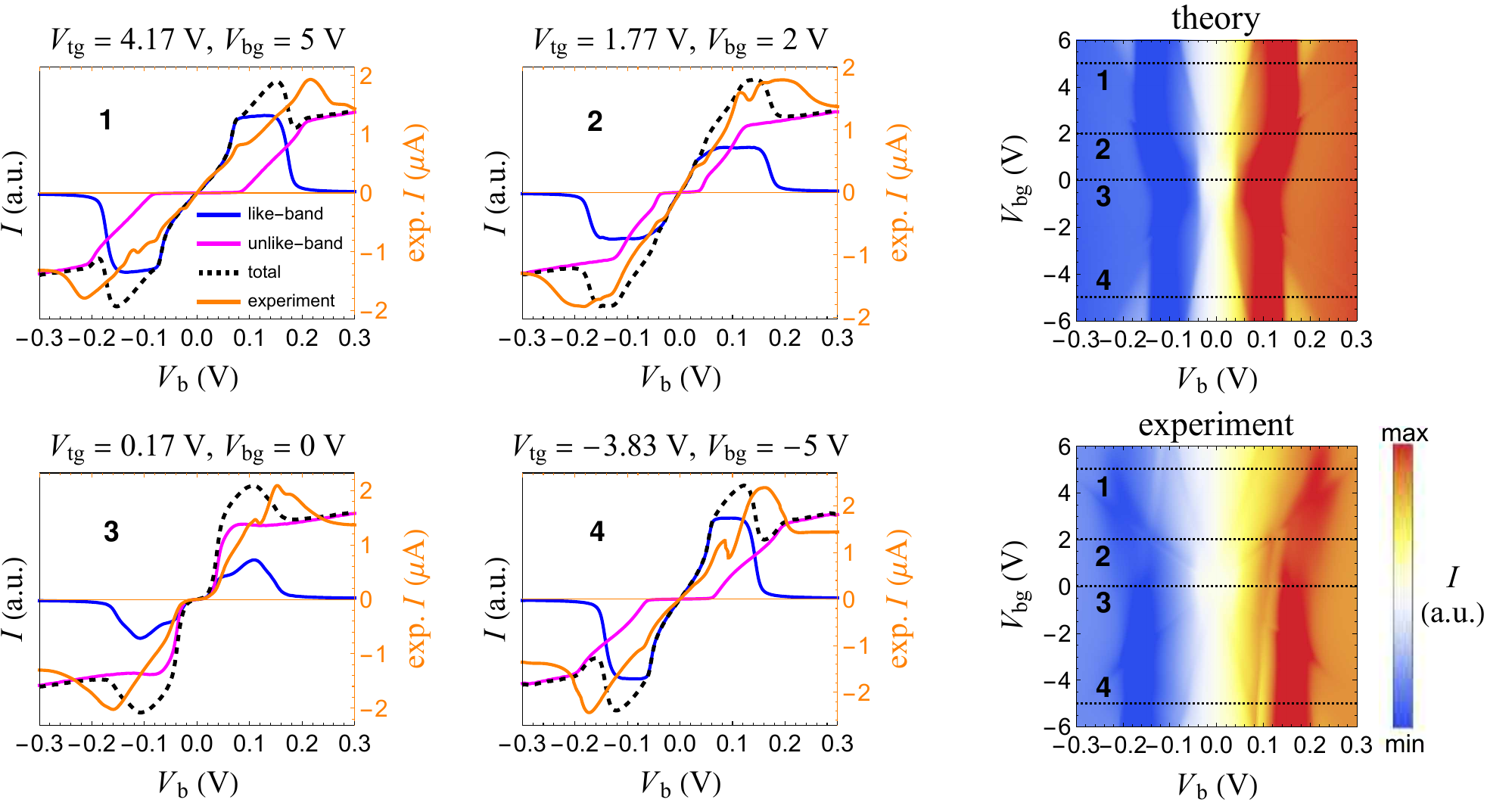}
\caption{Tunneling characteristics for Device 2 at different points $(V_\mathrm{tg},V_\mathrm{bg})$ of the equal-density diagonal. These characteristics are cross-sections (numbered by 1-4) of 2D tunneling current maps shown in right panels, which are the same as in Figure~\ref{FigST5f2}.}
\label{FigST5cf2}
\end{figure}

\begin{figure}[!p]         
\centering
\includegraphics[width=1\textwidth]{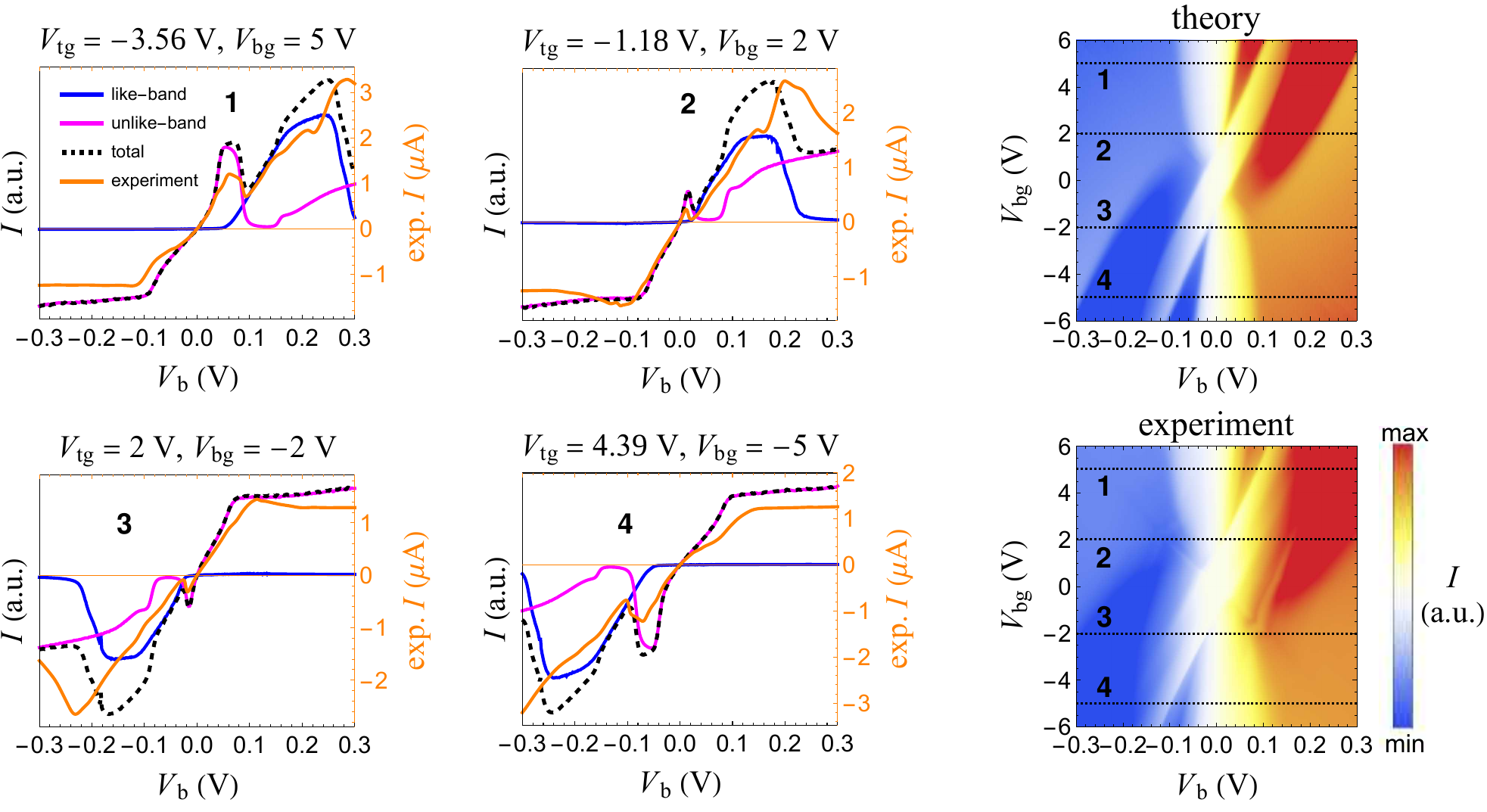}
\caption{The same as Figure~\ref{FigST5cf2}, but the points $(V_\mathrm{tg},V_\mathrm{bg})$ are taken at the opposite-density diagonal. 2D tunneling current maps in right panels are the same as in Figure~\ref{FigST5f1}.}
\label{FigST5cf1}
\end{figure}

\begin{figure}[!p]         
\centering
\includegraphics[width=1\textwidth]{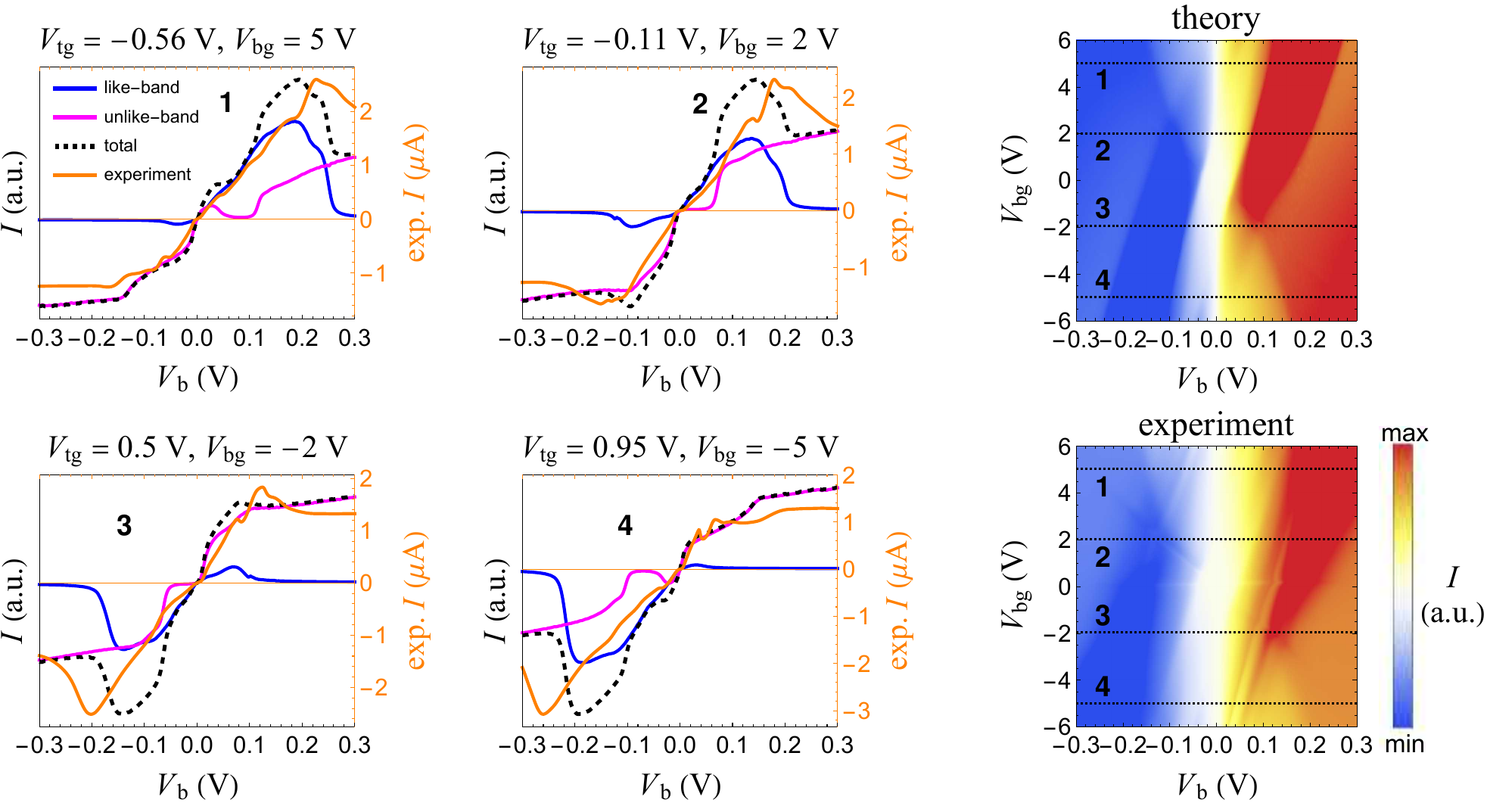}
\caption{The same as Figure~\ref{FigST5cf2}, but the points $(V_\mathrm{tg},V_\mathrm{bg})$ are taken to satisfy the charge neutrality condition of top BLG at $V_\mathrm{b}=0$. 2D tunneling current maps in right panels are the same as in Figure~\ref{FigST5f3}.}
\label{FigST5cf3}
\end{figure}

Here we present additional I-V characteristics for Device 1 (Figures~\ref{FigST4cf2}-\ref{FigST4cf5}) and Device 2 (Figures~\ref{FigST5cf2}-\ref{FigST5cf3}) taken at different combinations of gate voltages $(V_\mathrm{tg},V_\mathrm{bg})$. Those combinations which lie at the equal-density diagonal (Figures~\ref{FigST4cf2} and \ref{FigST5cf2}, and also point 3 in Figure~\ref{FigST4cf3}) demonstrate the symmetric I-V characteristics with resonant peaks of like-band current and steps of unlike-band current, as discussed in the main text.

In contrast, when the doping levels of two BLGs at $V_\mathrm{bg}=0$ are essentially different, i.e. when the point $(V_\mathrm{tg},V_\mathrm{bg})$ lies far from the equal-density diagonal, the I-V characteristics are asymmetric (Figures~\ref{FigST4cf1}-\ref{FigST4cf5} for Device 1 and \ref{FigST5cf1}-\ref{FigST5cf3} for Device 2). In this case the like-band current typically demonstrates the resonant peak when electron dispersions align, and the unlike-band current demonstrates the asymmetric peak caused by overlap of van Hove singularities, and additional steps far from the peaks.

The calculation generally agree with the experiments, reproducing all main qualitative features, although exact locations of the tunneling current maxima sometimes disagree due difficulties of approximating complicated electron dispersion in BLGs by our simple model. We also present in Figures~\ref{FigST4cf2}-\ref{FigST5cf3} absolute values of the measured tunneling current. Note that the current in Device 2 is by an order of magnitude larger than in Device 1 because of the multiply larger area of the device (overlap of BLG flakes).

\begin{figure}[!p]         
\centering
\includegraphics[width=0.95\textwidth]{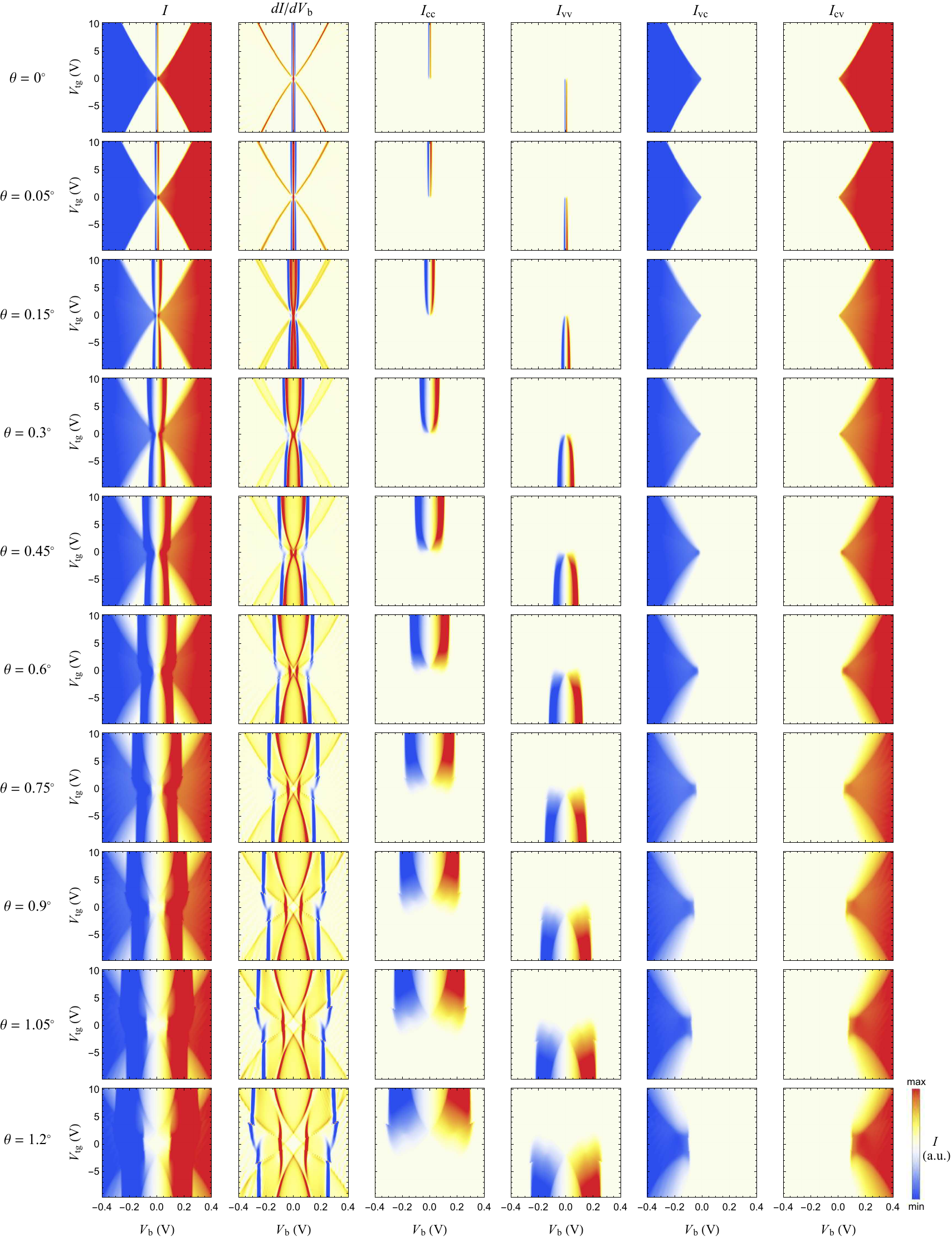}
\caption{Calculated 2D maps for tunneling current versus bias voltage $V_\mathrm{b}$ and $V_\mathrm{tg}$, when the latter changes simultaneously with $V_\mathrm{bg}$ to satisfy the condition of equal carrier densities at $V_\mathrm{b}=0$. Different rows correspond to successive increase of the twist angle $\theta$.}
\label{FigSAnglef2}
\end{figure}

\begin{figure}[!p]         
\centering
\includegraphics[width=0.95\textwidth]{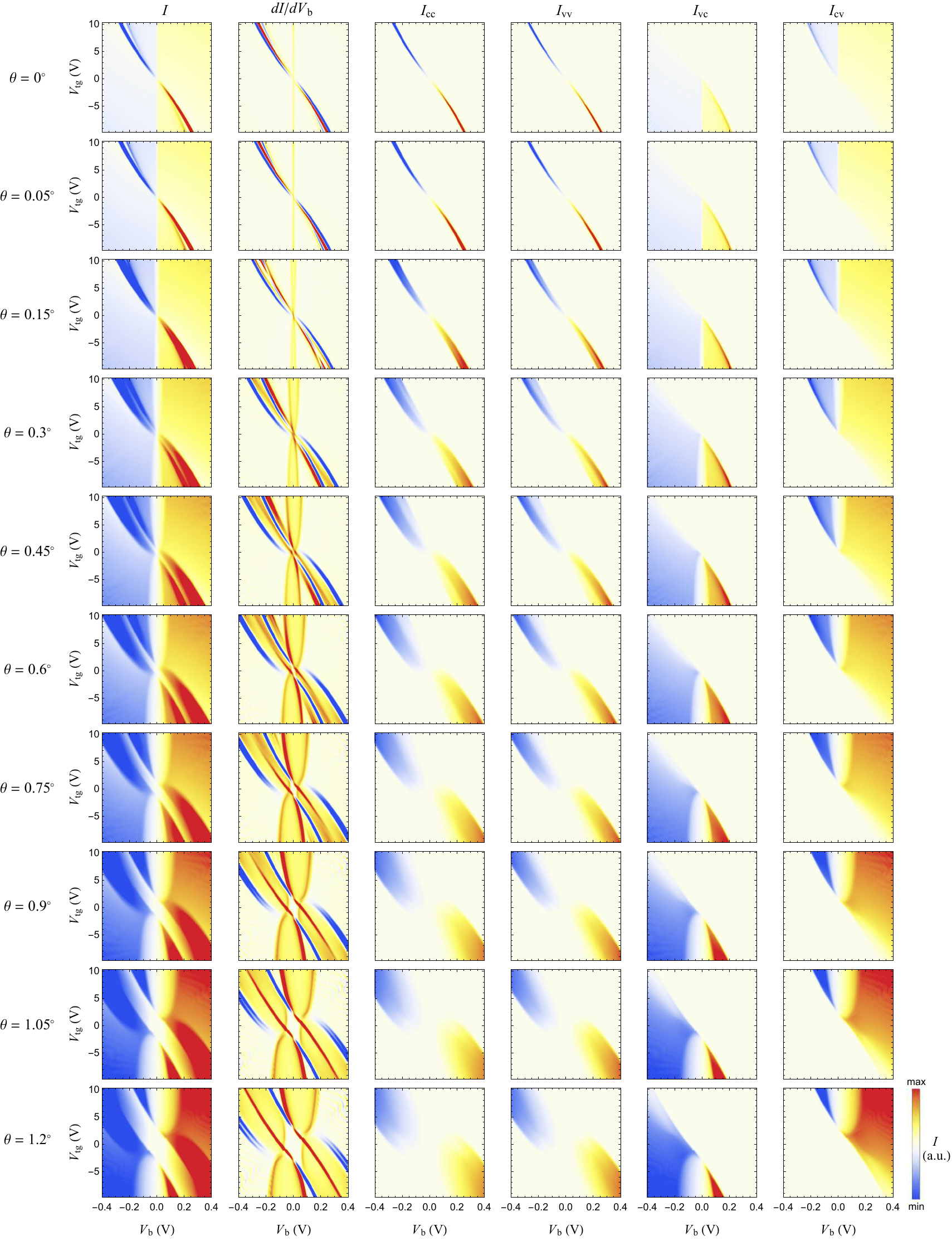}
\caption{Calculated 2D maps for tunneling current versus bias voltage $V_\mathrm{b}$ and $V_\mathrm{tg}$, when the latter changes simultaneously with $V_\mathrm{bg}$ to satisfy the condition of opposite carrier densities in both BLGs. Different rows correspond to successive increase of the twist angle $\theta$.}
\label{FigSAnglef1}
\end{figure}

\section{Twist angle progression}

In this section, we present the calculated 2D maps of the tunneling current along the equal-density (Figure~\ref{FigSAnglef2}) and opposite-density (Figure~\ref{FigSAnglef1}) diagonals at varying twist angle $\theta$. The calculations are carried out at the same parameters as for Device 1, where only $\theta$ was changed with keeping all other parameters the same.

The figures reveal how the tunneling maps typical to small twist angle (as for Device 1 with $\theta=0.13^\circ$) gradually transform becoming similar to the maps of Device 2 with $\theta=0.73^\circ$. As $\theta$ increases, the resonant peaks of the like-band currents ($I_\mathrm{cc}$ and $I_\mathrm{vv}$) become wider and shift to larger $|V_\mathrm{b}|$. The unlike-band currents ($I_\mathrm{cv}$ and $I_\mathrm{vc}$) demonstrate the different trend: the forbidden regions, where the current vanishes due to absence of interband overlaps, become wider at larger $\theta$, and the boundaries of the allowed regions become more diffuse, although the sublayer polarization of van Hove singularities makes the structure of these boundaries more complicated. The waves near the boundaries of some diagrams in Figures~\ref{FigSAnglef2} and \ref{FigSAnglef1} are insignificant numerical artifacts.

\printbibliography